\documentclass[prd,showpacs,preprintnumbers,
a4paper, nofootinbib ,twocolumn,floatfix
]{revtex4}
  
\usepackage{amsmath,amssymb}

\usepackage{epsfig}  
\usepackage{graphicx}               
\usepackage{url}

\newcommand{\nc}{\newcommand}  

\nc{\be}{\begin{equation}}  
\nc{\ee}{\end{equation}}  
\nc{\beqa}{\begin{eqnarray}}  
\nc{\eeqa}{\end{eqnarray}}  
\nc{\bea}{\begin{eqnarray}}  
\nc{\eea}{\end{eqnarray}}  
\nc{\ra}{\rightarrow}  
\nc{\lsim}{\begin{array}{c}\,\sim\vspace{-21pt}\\< \end{array}}  
\nc{\gsim}{\begin{array}{c}\sim\vspace{-21pt}\\> \end{array}}  
\nc{\cnone}{\tilde{\chi}^0_1}
\nc{\x}{{\bf x}}
\nc{\p}{{\cal P}}
\nc{\h}{{\cal H}}

\makeatletter
\def\url@leostyle{%
  \@ifundefined{selectfont}{\def\UrlFont{\sf}}{\def\UrlFont{\small\ttfamily}}}
\makeatother
\begin{document}
\preprint{FERMILAB-PUB-09-115-T}

\title{The PAMELA excess from neutralino annihilation in the NMSSM}

\author{Yang Bai${}^{a}$}
\author{Marcela Carena$^{a,b}$}
\author{Joseph Lykken$^{a}$}
\affiliation{  \vspace{3mm}
${}^{a}$Theoretical Physics Department, Fermilab, Batavia, Illinois 60510,
${}^{b}$Enrico Fermi Institute, Univ. of Chicago, 5640 Ellis Ave., Chicago, IL 60637}


   
\begin{abstract} 
We examine whether the cosmic ray positron excess observed by PAMELA can be explained by neutralino annihilation in the Next-to-Minimal Supersymmetric Standard Model (NMSSM). The main dark matter annihilation products are the lightest $CP$-even scalar $h_1$ plus the lightest $CP$-odd scalar $a_1$, with the $a_1$ decaying into two muons. The energetic positrons needed to explain PAMELA are thus obtained in the NMSSM simply from kinematics. The required large annihilation cross section is obtained from an $s$-channel resonance with the heavier $CP$-odd scalar $a_2$. Various experiments constrain the PAMELA-favored NMSSM parameter space, including collider searches for a light $a_1$. These constraints point to a unique corner of the NMSSM parameter space, having a lightest neutralino mass around 160 GeV and a very light pseudoscalar mass less than a GeV. A simple parameterized formula for the charge-dependent solar modulation effects reconciles the discrepancy between the PAMELA data and the estimated background at lower energies. We also discuss the electron and gamma ray spectra from the Fermi LAT observations, and point out the discrepancy   between the NMSSM predictions and Fermi LAT preliminary results and possible resolution. An NMSSM explanation of PAMELA makes three striking and uniquely correlated predictions: the rise in the PAMELA positron spectrum will turn over at around 70~GeV, the dark matter particle mass is less than the top quark mass, and a light sub-GeV pseudoscalar will be discovered at colliders.
\end{abstract}  

\maketitle

\section{Introduction} 
\setcounter{equation}{0}
Recently, the PAMELA collaboration has observed an anomalous positron abundance in cosmic radiation~\cite{Adriani:2008zr}. 
The positron over electron fraction turns over and appears to rise at energies from 10~GeV to 100~GeV. 
However, from the same detector, 
no obvious antiproton excess  is seen for the same energy range~\cite{Adriani:2008zq}. Many suggestions have been made to 
explain the positron excess at PAMELA. Among different approaches, dark matter annihilation is especially interesting and 
could imply future signals in dark matter direct detection experiments and/or at the Large Hadron Collider (LHC). 

In recent model-independent studies, dark matter particles are required both to annihilate dominantly to leptons 
and to have a much larger annihilation rate in the galactic halo than would be implied by a
traditional thermal relic estimate~\cite{Cirelli:2008jk}~\cite{Cirelli:2008pk}. 
The lack of antiproton excess in the PAMELA experiment~\cite{Adriani:2008zq} can be explained if dark matter particles 
first annihilate into some intermediate particles, which are so light that their decays to hadrons are kinematically 
forbidden~\cite{ArkaniHamed:2008qn}, or couple dominantly to Standard Model (SM) leptons~\cite{bai_fox}. 
To explain the large dark matter annihilation cross section in the galactic halo while remaining consistent with 
a thermal dark matter relic abundance, one can introduce an attractive force between two dark matter particles and use 
the Sommerfeld  enhancement to boost the annihilation cross section in the galactic halo~\cite{Hisano:2003ec}. 
This scenario suggests interesting signatures~\cite{ArkaniHamed:2008qp} and can be tested at the LHC~\cite{Bai:2009it}. 
A second approach is to consider non-thermal relics. Other long-lived particles can decay to 
the dark matter particles and increase the dark matter relic abundance in the galactic halo~\cite{Moroi:1999zb}. 

Instead of constructing a dark matter model \textit{ad hoc} to explain the cosmic ray observations, in this paper we ask whether 
there is an existing well-motivated model that naturally contains the necessary ingredients to explain the positron 
excess of PAMELA. The most developed framework to address the naturalness problem of the SM is the 
Minimal Supersymmetric Standard Model (MSSM), which provides the lightest superpartner (LSP) protected by 
R-parity as the dark matter candidate. In general, the annihilation products in the MSSM contain not only leptons 
but also a large fraction of hadrons. This makes it difficult for the MSSM to explain the positron excess at PAMELA, 
and at the same time be consistent with the antiproton spectrum at PAMELA~\cite{Grajek:2008pg}. From the theoretic side, 
the MSSM suffers the  $\mu$-problem, which can be solved elegantly by introducing a new gauge singlet chiral supermultiplet, 
as proposed in the Next-to-Minimal Supersymmetric Standard Model~\cite{Ellis:1988er}. A recent exploration of the 
NMSSM parameter space shows that the lightest $CP$-odd particle $a_1$ (mainly from the singlet component) can be naturally 
lighter than $2\,m_b$ and mainly decay into two $\tau$'s~\cite{Dermisek:2005ar}. One should notice that the mass of $a_1$ 
in the NMSSM is protected by the $U(1)_R$ symmetry and can even be lighter than $1$~GeV if the soft terms associated with 
the singlet are small.  In this case the $a_1$ will decay mainly  into two muons for a mass of a few hundred MeV. 

Therefore, if the dark matter candidate in the NMSSM can annihilate mostly into $a_1$'s, we can have a leptonic final state 
in the annihilation products simply from kinematics. The NMSSM provides the necessary ingredients to make this 
happen. Notice that the dark matter candidate in the NMSSM is the LSP neutralino (for existing studies for light neutralino 
dark matter in the NMSSM, see~\cite{Gunion:2005rw}).  If its mass happens to be around half that of the heavier $CP$-odd 
scalar $a_2$ mass,  a large dark matter annihilation cross section is obtained through the $s$-channel resonance 
effects with $a_2$. To have leptons dominant in the final state, we should have a large branching ratio of $a_2$ to $a_1$ plus $h_1$. 
This will naturally happen, provided that the dark matter LSP mass is less than the top quark mass, and thus that the decay of $a_2$ to $t\,\bar{t}$ 
is kinematically forbidden. Since the $CP$-odd scalar coupling to other fermions is proportional to their Yukawa couplings, 
the final state can dominantly be $a_1\,+\,h_1$ with the former decaying to leptons. The kinematics helps us to obtain hard 
leptons in the final state of the dark matter annihilations. 

In Section~\ref{sec:model}, we develop the notation by deriving the spectrum and 
interactions in the NMSSM, and show two sets of representative model points allowed by current experimental constraints. 
We calculate the dark matter annihilation cross section in Section~\ref{sec: annihilation} and positron excesses from 
neutralino annihilation in Section~\ref{sec:positrons}. In Section~\ref{sec:pamela-proton}, we consider the constraints 
on the model parameter space from the PAMELA antiproton spectrum. We discuss the gamma-ray spectrum in Section~\ref{sec:gammaray} 
and point out a discrepancy with recent preliminary Fermi LAT results and comment on a possible resolution. In Section~\ref{sec:constraints} we demonstrate the consistency of our model points with
direct constraints from LEP, Tevatron, CLEO, B-factories and the magnetic 
moment of the muon. Finally, we discuss dark matter direct detection and conclude  
in Section~\ref{sec:conclusions}.

\section{Spectrum and interactions in the NMSSM}
\label{sec:model}
To describe the NMSSM model, we follow the notation in the Ref.~\cite{Ellwanger:2004xm}.
The superpotential in the NMSSM is 
\beqa
W\,=\,\lambda\,\hat{S}\,\hat{H_u}\,\hat{H_d}\,+\,\frac{\kappa}{3}\,\hat{S}^3\,,
\eeqa
and the soft supersymmetry-breaking terms are 
\beqa
V\,=\,\lambda\,A_\lambda\,S\,H_u\,H_d\,+\,\frac{\kappa}{3}\,A_\kappa\,S^3\,+\,h.c.\,.
\eeqa
Here, hatted capital letters denote superfields, and unhatted capital letters the corresponding scalar components. 
The minimization of the scalar potential determines their vacuum expectation values (VEVs): $h_u\equiv \langle H_u\rangle$, 
$h_d\equiv \langle H_d\rangle$ and $s\equiv \langle S\rangle$. The electroweak scale $v=\sqrt{h_u^2\,+\,h_d^2}=174$~GeV. 
Since $\mu_{\rm eff}=\lambda\,s$, there are four new parameters in the NMSSM,
which we take to be real: $\lambda$, $A_\lambda$, 
$\kappa$ and $A_\kappa$. With sign conventions for the fields, $\lambda$ and $\tan{\beta}\equiv h_u/h_d$ are positive, 
while $A_\lambda$, $\kappa$, $A_\kappa$ and $\mu_{\rm eff}$ can have either sign. 

There exists a ${\cal Z}_3$ symmetry for the NMSSM, which is spontaneously broken and induces a domain wall problem. 
One can introduce higher dimension operators to explicitly break this discrete symmetry and perhaps
circumvent this problem~\cite{Panagiotakopoulos:1998yw}. Since those ${\cal Z}_3$ breaking operators have small 
effects on the analysis performed in this paper, we will neglect them from now on.

\subsection{Neutralinos}
\label{sec:neutralino}
There are five neutralinos in the NMSSM: the $U(1)_Y$ gaugino $\lambda_1$, the neutral $SU(2)_W$ gaugino $\lambda_2$, 
the Higgsinos $\psi^0_u$ and  $\psi^0_d$ and the singlino $\psi_s$. In the basis $\psi^0=(-i\lambda_1, -i\lambda_2 ,  
\psi^0_u, \psi^0_d  , \psi_s    )$, we have the neutralino mass matrix
\beqa
{\cal L}\,=\,-\frac{1}{2}\,(\psi^0)^T\,{\cal M}_0\,(\psi^0)\,+\,h.c.\,,
\eeqa
where
\beqa
{\cal M}_0=\left( \begin{array}{ccccc}
M_1 & 0 & \frac{g_1\,h_u}{\sqrt{2}} & -\frac{g_1\,h_d}{\sqrt{2}} & 0 \vspace{1mm} \\
0     & M_2 & -\frac{g_2\,h_u}{\sqrt{2}} & \frac{g_2\,h_d}{\sqrt{2}} & 0 \vspace{1mm} \\
 \frac{g_1\,h_u}{\sqrt{2}} &  -\frac{g_2\,h_u}{\sqrt{2}}  &  0 & -\mu & -\lambda\, h_d \vspace{1mm}  \\
 -\frac{g_1\,h_d}{\sqrt{2}} & \frac{g_2\,h_d}{\sqrt{2}}  &  -\mu & 0 &  -\lambda\, h_u  \vspace{1mm} \\
 0 &     0   &     -\lambda\, h_d    & -\lambda\, h_u    &    2\,\kappa\,s
\end{array}  \right).
\eeqa
To obtain the needed dark matter annihilation rate,
we consider the case that the LSP, the lightest neutralino, is mainly made of the bino with  
mixings with Higgsinos. For $g_i\,v,\lambda\,v \ll  |M_1| < |\mu| < |M_2|$, a moderate $\tan{\beta}>1$ 
and $\mu < 0$, the lightest neutralino is approximately 
\beqa
\chi&\approx& -i\lambda_1\,-\,(\sin{\alpha_1}\,\cos{\beta}\,-\,\sin{\alpha_2}\,\sin{\beta})\,\psi^0_u   \nonumber \\
&&
\,-\,(\sin{\alpha_2}\,\cos{\beta}\,+\,\sin{\alpha_1}\,\sin{\beta})\,\psi^0_d\,,
\eeqa
with
\beqa
\alpha_1&=&\frac{1}{2}\,\arctan{\frac{\sqrt{2}\,g_1\,v\,\mu\,\cos{2\beta}}{\mu^2\,-\,M_1^2\,+\,(\lambda^2\,-\,g_1^2/2)v^2}}  \,,\label{eq:alpha1} \\
\alpha_2&=&\frac{1}{2}\,\arctan{\frac{\sqrt{2}\,g_1\,v\,(M_1\,+\,\mu\,\sin{2\beta})}{M_1^2\,-\,\mu^2}}    \,,\label{eq:alpha2}
\eeqa
and with its mass approximated as
\beqa
m_{\chi}\,=\,M_1\,+\,\frac{g_1^2\,v^2(M_1\,+\,\mu\,\sin{2\beta})}{2\,(M_1^2\,-\,\mu^2)}  \,.
\eeqa
Hereafter, we use a simple notation $\chi$ to replace the usual notation $\cnone$ for the lightest neutralino. 
\subsection{Higgs sector at tree level}
\label{sec:higgs}
The charged Higgs $H^{\pm}\,=\,\cos{\beta}\,H_u^\pm\,+\,\sin{\beta}\,H_d^\pm$ has a mass
\beqa
M^2_{H^\pm}\,=\,\lambda\,s\,(A_\lambda\,+\,\kappa\,s)\,\frac{2}{\sin{2\beta}}\,+\,(\frac{g^2_2}{2}\,-\lambda^2)\,v^2\,.
\eeqa
For $\kappa$ and $\lambda$ of order of unity, the mass of $H^\pm$ is generically ${\cal O}(\mu)$.

Expanding around the Higgs fields VEVs, the neutral scalar fields are defined as
\beqa
H^0_u&=&h_u\,+\,\frac{H_{uR}\,+\,i\,H_{uI}}{\sqrt{2}}\,,\nonumber \\
 H^0_d&=&h_d\,+\,\frac{H_{dR}\,+\,i\,H_{dI}}{\sqrt{2}}\,, \nonumber \\
  S&=&s\,+\,\frac{S_R\,+\,i\,S_I}{\sqrt{2}}\,.
\eeqa
For the three $CP$-even neutral states, we can diagonalize their $3\times 3$ mass matrix by an orthogonal matrix 
$S_{ij}$ to obtain the mass eigenstates (ordered in mass): $h_i\,=\,S_{ij}\,(H_{uR}, H_{dR}, S_R)_j$\,, 
with masses denoted by $m_{h_i}$\,. For $v\,\ll\,s$, we write the lightest $CP$-even Higgs as
\beqa
h_1&\approx&\cos{\alpha}\,\cos{\theta_S}\,H_{uR}\,+\,\sin{\alpha}\,\cos{\theta_S}\,H_{dR} \nonumber \\
&&\,-\,\sin{\theta_S}\,S_R\,,\label{eq:h1}
\eeqa
with 
\beqa
\alpha&\approx&\frac{\pi}{2}\,-\,\beta\,-\,\frac{\lambda\,M_Z^2\,\sin{2\beta}\,\sin{4\beta}}{4\,\kappa\,\mu^2}\,, \nonumber \\
\theta_S&=&-\,\frac{\lambda^2\,v}{2\kappa^2\,s}\,\left(1\,-\,\frac{\kappa}{\lambda}\,\sin{2\beta}\right)\,+\,{\cal O}(v^3/s^3)\,.
\eeqa
The singlet component of $h_1$ is small and suppressed by $v/s$.

There are three $CP$-odd pseudoscalar fields, one of which is a massless 
Goldstone mode eaten by the $Z$ boson. Dropping the Goldstone mode, 
the remaining $2\times 2$ mass matrix in the $(\tilde{A}, S_I)$ basis with $\tilde{A}\equiv \cos{\beta}\,H_{uI}\,+\,\sin{\beta}\,H_{dI}$\,, is 
\begin{widetext}
\beqa
{\cal M}_{\rm odd}\,=\,\left(
\begin{array}{cc}
\lambda\,s\,\frac{h_u^2\,+\,h_d^2}{h_u\,h_d}(A_\lambda\,+\,\kappa\,s) &  \lambda\,\sqrt{h_u^2\,+\,h_d^2}\,(A_\lambda\,-\,2\,\kappa\,s) \vspace{3mm}  \\
\lambda\,\sqrt{h_u^2\,+\,h_d^2}\,(A_\lambda\,-\,2\,\kappa\,s) & 
 4\,\lambda\,\kappa\,h_u\,h_d\,+\,\lambda\,A_\lambda\,\frac{h_u\,h_d}{s}\,-\,3\,k\,A_\kappa\,s
\end{array}
\right) \,.
\eeqa
We introduce a mixing angle $\theta_A$ to diagonalize the above matrix:
\beqa
\tan{\theta_A}=-\,\frac{2\,s\,(A_\lambda\,+\,\kappa\,s)}{v\,(A_\lambda\,-\,2\,\kappa\,s)\,\sin{2\beta}}\, ,\qquad
\cos^2{\theta_A}=\frac{v^2\,(A_\lambda\,-\,2\,\kappa\,s)^2\,\sin^2{2\beta}}{4\,s^2\,(A_\lambda\,+\,\kappa\,s)^2\,
+\,v^2\,(A_\lambda\,-\,2\,\kappa\,s)^2\,\sin^2{2\beta}}\,,
\eeqa
\end{widetext}
and arrive at the physical $CP$-odd states $a_i$ (ordered in mass) 
\beqa
a_1&=&\cos{\theta_A} (\cos{\beta}\,H_{uI}\,+\,\sin{\beta}\,H_{dI})\,+\,\sin{\theta_A}\,S_I \,, \nonumber \\
a_2&=&-\sin{\theta_A} (\cos{\beta}\,H_{uI}\,+\,\sin{\beta}\,H_{dI})\,+\,\cos{\theta_A}\,S_I \,.
\nonumber \\
\eeqa
The lightest $CP$-odd particle $a_1$ is mainly composed of the singlet field when $\cos{\theta_A}\rightarrow 0$, and is a doublet field otherwise. 
Since we are looking for a very light scalar, we observe from the determinant of ${\cal M}_{\rm odd}$ that this occurs if
$A_\kappa$ and $A_\lambda$ are small. This is technically natural since $A_\kappa,A_\lambda \to 0$ is a symmetry-enhancing limit. 

The heavier $CP$-odd scalar mass is approximately 
\beqa
M^2_{a_2}&=&\frac{2\,\lambda\,s\,(A_\lambda\,+\,\kappa\,s)}{\sin{2\beta}}  \nonumber \\
&&\quad\,+\,\frac{\lambda\,v^2\,(A_\lambda\,-\,2\,\kappa\,s)^2\,\sin{2\beta}}{2\,s\,(A_\lambda\,+\,\kappa\,s)}\,.
\eeqa
with the condition $A_\lambda\,+\,\kappa\,s \,>\, 0$. 

For small $A_\lambda$ and $A_\kappa$, we have the following approximate formulae:
\beqa
M^2_{a_1}&\approx&\frac{3\,s\,(3\,\lambda\,A_\lambda\,v^2\,\sin{2\beta}\,-\,2\,\kappa\,A_\kappa\,s^2)}{2(s^2\,+\,v^2\,\sin^2{2\beta})}\,,  \label{eq:ma1}  \\
M^2_{a_2}&\approx&\frac{2\,\kappa\,\lambda\,s^2}{\sin{2\beta}}\,+\,2\,\kappa\,\lambda\,v^2\,\sin{2\beta}  \,,   \label{eq:ma2}
\eeqa
and 
\beqa
\tan{\theta_A}&\approx&\frac{s}{v\,\sin{2\beta}}\,,  \nonumber \\
\cos^2{\theta_A}&\approx&\frac{v^2\,\sin^2{2\,\beta}}{v^2\,\sin^2{2\,\beta}\,+\,s^2}
\,\approx\, \frac{v^2\,\sin^2{2\,\beta}}{s^2} 
\,,
\eeqa
for $v\ll s$.

\subsection{Interactions and decay modes}
\label{sec:interactions}
Since we are interested in dark matter annihilation through $s$-channel $a_2$ exchange, 
we list the relevant vertices associated with $CP$-odd scalars in this section. 
For $v \ll s$, the couplings of $CP$-odd scalars to fermions are 
\beqa
&&a_2\,t_L\,t^c_R\,: \; -\,i\,\frac{m_t\,\sin{\theta_A}}{\sqrt{2}\,v\,\tan{\beta}}\,, \nonumber \\
&&a_2\,b_L\,b^c_R\,: \; i\,\frac{m_b\,\tan{\beta}\,\sin{\theta_A}}{\sqrt{2}\,v}\,,  \nonumber \\
&&a_1\,t_L\,t^c_R\,: \; -\,i\,\frac{m_t\,\cos{\theta_A}}{\sqrt{2}\,v\,\tan{\beta}}\,, \nonumber \\
&&a_1\,b_L\,b^c_R\,: \; i\,\frac{m_b\,\tan{\beta}\,\cos{\theta_A}}{\sqrt{2}\,v}\,.
\eeqa
The couplings of $a_2$ and $a_1$ to the lightest neutralino are
\beqa
 i\,g_{a_2\chi\chi}&\equiv&i\,g_1\,\sin{\theta_A}\,(\cos{2\beta}\,\sin{\alpha_1}\,-\,\sin{2\beta}\,\sin{\alpha_2})   \,, \nonumber \\ \label{eq:a2chichi} \\
 i\,g_{a_1\chi\chi}&\equiv&-\,i\,g_1\,\cos{\theta_A}\,(\cos{2\beta}\,\sin{\alpha_1}\,-\,\sin{2\beta}\,\sin{\alpha_2})   \,, \nonumber \\ \label{eq:a1chichi}
\eeqa
with $\alpha_1$ and $\alpha_2$ defined in Eqs.~(\ref{eq:alpha1}--\ref{eq:alpha2}). Therefore, 
the coupling of $a_2\,\chi\,\chi$ increases as one increases the Higgsino components of the LSP. Their couplings to gauge bosons are
\begin{widetext}
\beqa
&&\hspace*{-20pt}
a_2(p)\,H^+(p^\prime)\,W^-_\mu:\; -\,\frac{i\,g_2\,\sin{\theta_A}}{2}\,(p\,-\,p^\prime)_\mu \,, \\
&&\hspace*{-20pt}
a_1(p)\,H^+(p^\prime)\,W^-_\mu:\; \,\frac{i\,g_2\,\cos{\theta_A}}{2}\,(p\,-\,p^\prime)_\mu \,, \\
&&\hspace*{-20pt}
a_2(p)\,h_1(p^\prime)\,Z_\mu:\; -\frac{i\,g}{\sqrt{2}}\sin{\theta_A}\cos{(\alpha+\beta)}(p^\prime-p)_\mu 
\approx - \frac{i\,g}{\sqrt{2}}\frac{M_Z^2\,\sin{\theta_A}\,\sin{4\beta}}{2\,M_{a_2}^2}\,(p^\prime - p)_\mu \,, \label{eq:a2h1Z} \\
&&\hspace*{-20pt}
a_1(p)\,h_1(p^\prime)\,Z_\mu:\; \frac{i\,g}{\sqrt{2}}\,\cos{\theta_A}\,\cos{(\alpha+\beta)}\,(p^\prime\,-\,p)_\mu 
\approx\frac{i\,g}{\sqrt{2}}\,\frac{M_Z^2\,\cos{\theta_A}\,\sin{4\beta}}{2\,M_{a_2}^2}\,(p^\prime\,-\,p)_\mu \,. \label{eq:a1h1Z} 
\eeqa
\end{widetext}
The couplings among $a_2$, $a_1$ and $h_1$ depend on the diagonalization of the $CP$-even scalar mass matrix, 
which may have significant one-loop contributions. For simplicity, we use the tree-level results in Eq.~(\ref{eq:h1}) 
to obtain analytic formulae. We arrive at the following dimensional couplings to the leading power in $v/s$~\cite{Dermisek:2006wr}
\beqa
a_1\,a_1\,h_1\,: && w_{a_1a_1h_1}= {\cal O}(v^3/s^2)\,,   \\
a_2\,a_1\,h_1\,: && w_{a_2a_1h_1}=-\sqrt{2}\kappa\mu+{\cal O}(\lambda^2\,v^2/s)\,.
\eeqa
Here the approximation is valid for small values of $A_\lambda$ and $A_\kappa$. 
We also need the main 
decay channels of $h_1$ and $a_2$. For the $CP$-even particle $h_1$ with a mass below $2\,M_W$, 
it mainly decays into 2 $a_1$'s or 2 $b$'s with the decay widths calculated as following:
\beqa
\Gamma(h_1\,\rightarrow\,2\,a_1)&= &\frac{1}{32\,\pi\,M_{h_1}} \,{\cal O}(v^6/s^4)              \,,\\
\Gamma(h_1\,\rightarrow\,b\,+\,\bar{b})&\approx&\frac{3\,M_{h_1}}{8\,\pi}  \, \left(\frac{m_b}{\sqrt{2}\,v}\right)^2\,.
\eeqa
Thus for $v/s\ll 1$ the decay $h_1\rightarrow 2\,a_1$ is suppressed in favor of
$h_1 \rightarrow b\,\bar{b}$. Note this suppression is directly connected to the
small values of $A_\lambda$ and $A_\kappa$. For large values of $A_\lambda$ and $A_\kappa$, 
the $h_1\rightarrow 2\,a_1$ decay would be the dominant one~\cite{Dermisek:2006wr}.

Similarly for $a_2$, if its mass is below twice the top quark mass, the leading two decay channels are 
\beqa
\Gamma(a_2\,\rightarrow\,h_1\,a_1)&\approx& \frac{\kappa\,\lambda}{32\,\pi} \, M_{a_2}\,\sin{2\beta}     \,,\\
\Gamma(a_2\,\rightarrow\,b\,+\,\bar{b})&\approx&\frac{3\,M_{a_2}}{8\,\pi}  \, \left(\frac{m_b\,\tan{\beta}\,\sin{\theta_A}}{\sqrt{2}\,v}\right)^2\,.
\eeqa
The bosonic decay channel can be 
dominant for modest values for $\kappa$ and $\lambda$. However, if the $a_2$ mass exceeds
twice the top quark mass, the decay channel into $t\,\bar{t}$ opens:
\beqa
\Gamma(a_2\,\rightarrow\,t\,+\,\bar{t})&\approx&\frac{3\,M_{a_2}}{8\,\pi}  \, \left(\frac{m_t\,\sin{\theta_A}}{\sqrt{2}\,v\,\tan{\beta}}\right)^2 \nonumber \\
&& \times\,  \sqrt{1\,-\,\frac{4\,m_t^2}{M_{a_2}^2}}      \,,
\eeqa
and this would become the dominant decay for $a_2$ assuming $\kappa\,,\lambda <1$.

\subsection{Spectrum from numerical calculations}
\label{sec:decay}
In this section we find NMSSM model points that can provide the neutralino as a 
DM candidate to explain PAMELA. There are two relevant possibilities depending on the mass of $a_1$:
\begin{enumerate}
 \item The mass of the lightest $CP$-odd particle $a_1$ is in the range $(2\,m_\mu,  1~{\rm GeV})$\,.
  \item The mass of the lightest $CP$-odd particle $a_1$ is in the range $(2\,m_\tau, 2\,m_b)$\,.
 \end{enumerate}
In the first case $a_1$ mainly decays to two muons because decays to mesons are 
kinematically suppressed. Because $a_1$ is $CP$-odd, decays to 
two pions are forbidden due to CP symmetry, while decays to three pions are suppressed by the 
three-body phase space. In the following, we will approximate $a_1\rightarrow 2\mu$ as 100\% 
for case 1. In the second case, because the couplings of $a_1$ to fermions are proportional 
to fermion masses, we anticipate that $a_1$ decays mainly to 2 $\tau$'s. We do not consider 
the case that $a_1$ mainly decays into two electrons because of stringent constraints from 
the beam-dump experiment at CERN~\cite{Bergsma:1985qz}. 
 
To find the interesting parts of the parameter space, we use the program NMHDECAY~\cite{Ellwanger:2004xm}
for numerical checks. There are many experimental constraints considered in NMHDECAY 
such as  various Higgs searches at LEP, $b\rightarrow s\,\gamma$ and $\Upsilon(1S)\rightarrow a_1\,\gamma$. 
The model points presented in this paper pass all of the constraints embedded in NMHDECAY. 
Furthermore, we will discuss updated constraints on $\Upsilon(3S) \rightarrow a_1\,\gamma \rightarrow \mu^+\,\mu^-\,\gamma$ 
from BaBar and searches for dimuon resonances at  
LEP and Tevatron in section~\ref{sec:constraints}.
%
\begin{widetext}
\begin{table}[htb]
\renewcommand{\arraystretch}{1.6}
\begin{center}
\begin{tabular*}{1.0\textwidth} {@{\extracolsep{\fill}} cccccccc} 
\hline \hline
$\tan{\beta}$ & $\lambda$  & $\kappa$ & $A_\lambda$   &    $A_\kappa$   &  $\mu_{\rm eff}$   & $M_1$ & $M_2$  \\ \hline 
3.1  & 0.24    & 0.194 &  -0.05 & -0.273 & -190 & 178.5 & 200 \\
\hline \hline
$m_\chi$ &  $M_{a_1}$ &  $M_{a_2}$    &   $M_{h_1}$    &  $M_{h_2}$       &  $M_{H^\pm}$     &     $m_{\chi^\pm}$  & $\Gamma_{a_2}$   \\
\hline
161.8 &   0.81   & 320.1  &   114.7   & 297.6     &  325.4   &  175.6   &  0.22   \\
\hline \hline
   \multicolumn{2}{c}{${\rm Br}(h_1\rightarrow b \bar{b})=78.3\%$}    & &  \multicolumn{2}{c}{${\rm Br}(h_1\rightarrow \tau \bar{\tau})=8.1\%$}  &  \multicolumn{2}{c}{${\rm Br}(h_1\rightarrow a_1 a_1)=1.0\%$} &      \\
   \hline
      \multicolumn{2}{c}{${\rm Br}(a_1\rightarrow \mu^+ \mu^-)\approx 100\%$}  &   &  &  &   &    \\   
  \hline
      \multicolumn{2}{c}{${\rm Br}(a_2\rightarrow a_1 h_1)= 68.0\%$}  &   &  \multicolumn{2}{c}{${\rm Br}(a_2\rightarrow b \bar{b})= 22.5\%$}   &     \multicolumn{2}{c}{${\rm Br}(a_2\rightarrow Z h_1 )=5.4\%$} &  \\  
      \hline \hline
  \multicolumn{6}{c}{$\chi\,=\,-0.595\,(-i\lambda_1)\,+\,0.347\,(-i\lambda_2)\,+\,0.599\,( \psi^0_u)\,+\,0.404\,( \psi^0_d)\,-\,0.061\,(\psi_s  )$} &    \multicolumn{2}{c}{$\cos{\theta_A}=0.12$}\\
\hline \hline
\end{tabular*}
\vspace{2mm}
\caption{$\mu$-favored model point. Masses are in GeV.}
\label{tab:case1para}
\end{center}
\end{table}
\end{widetext}

For both cases, in order to isolate the dark matter discussion, we choose the less relevant 
soft terms to be heavy. For example, we choose 500~GeV soft masses for sleptons, 1~TeV for squarks, 1~TeV for 
gluino and $-2.5$~TeV for all the $A$-terms in the quark and lepton sectors.
We have used the updated top quark mass $m_t\,=\,173.1\pm 0.6 \pm 1.1$~GeV~\cite{topquarkmass}, 
which has a significant correlation to the Higgs boson mass for small $\tan{\beta}$ as considered in this paper. 

For the muon-favored case, we choose values for other parameters in the NMSSM as in Table~\ref{tab:case1para}, which also shows the relevant spectrum and branching ratios of the light scalars. The default lower limit on the $a_1$ mass is 1~GeV in NMHDECAY. 
One has to change the code file named {\tt mhiggs.f} to obtain an $a_1$ mass below 1~GeV. As can be seen from Table~\ref{tab:case1para}, if $a_2$ can be produced from $\chi \chi$ annihilation 
through the resonance effect, the final products of the DM annihilation mainly contain $a_1+ h_1$. 
The $a_1$ decays into two muons to provide the positrons needed to explain the excess at PAMELA. 
The dark matter mass is mainly controlled by the parameter $M_1$ in this model, which is chosen to have the 
lightest neutralino mass below the top quark mass. Otherwise, $a_2$ will decay into $t\,\bar{t}$ with a 
significant branching ratio, and the neutralino annihilation produces a limited amount of positrons and 
lots of hadrons, the hadrons being disfavored by the null 
antiproton excess at PAMELA. It is intriguing that a combination of NMSSM and PAMELA results forces us to 
have a dark matter mass below around 170~GeV.  

It is technically natural to have  $M_{a_1}$ 
below 1~GeV for tiny values of $A_\lambda$ and $A_\kappa$ as reported here, since the $U(1)_R$ symmetry protects 
its mass (one can also use the $U(1)_{PQ}$ symmetry to obtain a light pseudoscalar, see~\cite{Thaler} for example). 
Notice that the branching ratio of $h_1\rightarrow a_1a_1$ is below 1.2\% to satisfy the current 
null results of searches of $a_1$  in the channel $h_1\rightarrow a_1a_1 \rightarrow 4\,\mu$  at D0 (see Section~\ref{sec:TEVATRON}).
 
For the tau-favored case, we list the values of model parameters, spectrum of particles and 
interesting branching ratios of light scalars in Table~\ref{tab:case2para}. The current direct 
searches only impose mild constraints on the model parameters. Therefore, we choose one 
representative point in the parameter space to have $h_1\rightarrow a_1a_1 \rightarrow 4\tau$ 
as the main decay channel of $h_1$, and hence to have six $\tau$'s in the final state of dark matter annihilations. 
%
\begin{widetext}
\begin{table}[htb]
\renewcommand{\arraystretch}{1.6}
\begin{center}
\begin{tabular*}{1.0\textwidth} {@{\extracolsep{\fill}} c  c  c  c c  c  c  c } 
\hline \hline
$\tan{\beta}$ & $\lambda$  & $\kappa$ & $A_\lambda$   &    $A_\kappa$   &  $\mu_{\rm eff}$   & $M_1$ & $M_2$  \\ \hline 
2.0  & 0.519    & 0.458 &  -14.83 & -3.6 & -200.0 & 162.25 & 1000 \\
\hline \hline
$m_\chi$ &  $M_{a_1}$ &  $M_{a_2}$    &   $M_{h_1}$        &         $M_{h_3}$  &  $M_{H^\pm}$     &     $m_{\chi^\pm}$ & $\Gamma_{a_2}$    \\
\hline
161.7&   7.8   & 322.0  &   123.0  & 296.9     & 304.7    &   208.0  & 0.79   \\
\hline \hline
   \multicolumn{2}{c}{${\rm Br}(h_1\rightarrow a_1 a_1)=92.2\%$}    & &  \multicolumn{2}{c}{${\rm Br}(h_1\rightarrow b\bar{b})=7.6\%$}  & &      \\
   \hline
      \multicolumn{2}{c}{${\rm Br}(a_1\rightarrow \tau^+ \tau^-)=85.0\%$}  &   &  \multicolumn{2}{c}{${\rm Br}(a_1\rightarrow gg)=7.7\%$}   &   &  \multicolumn{2}{c}{${\rm Br}(a_1\rightarrow c\bar{c})=5.4\%$}   \\   
  \hline
      \multicolumn{2}{c}{${\rm Br}(a_2\rightarrow a_1 h_1)=95.5\%$}  &   &  \multicolumn{2}{c}{${\rm Br}(a_2\rightarrow b\bar{b})=2.8\%$}   &   &  \multicolumn{2}{c}{${\rm Br}(a_2\rightarrow a_1 h_2)=0.7\%$}   \\  
            \hline \hline
  \multicolumn{7}{c}{$\chi\,=\,0.968\,(-i\lambda_1)\,+\,0.003\,(-i\lambda_2)\,-\,0.236\,( \psi^0_u)\,-\,0.080\,( \psi^0_d)\,+\,0.043\,(\psi_s  )$} & \\
\hline \hline
\end{tabular*}
\vspace{2mm}
\caption{$\tau$-favored model point. The masses are in GeV.}
\label{tab:case2para}
\end{center}
\end{table}
\end{widetext}
%

\section{Dark matter annihilation cross section}
\label{sec: annihilation}
Our first goal is to look for parameter space in the NMSSM having leptons as the 
main dark matter annihilation products. This goal can be achieved if the lightest $CP$-odd scalar is 
an intermediate annihilation product and subsequently decays to two taus or two muons from 
kinematic constraints. The other goal is to have a large dark matter annihilation 
cross section to explain the size of the PAMELA excess. The annihilation cross section can be 
enhanced through $s$-channel resonance effects. Assuming the lightest neutralino is the dark matter 
candidate in the NMSSM, the heavier $CP$-odd scalar $a_2$ is the only particle which can play this role. 
The $CP$-even scalars are ruled out by the $CP$ symmetry, because the initial state with two identical 
Majorana fermions is $CP$-odd. The lighter $CP$-odd scalar $a_1$ is far below the necessary mass region 
for the resonance effect.

\begin{figure}[ht!]
\includegraphics[width=0.47\textwidth]{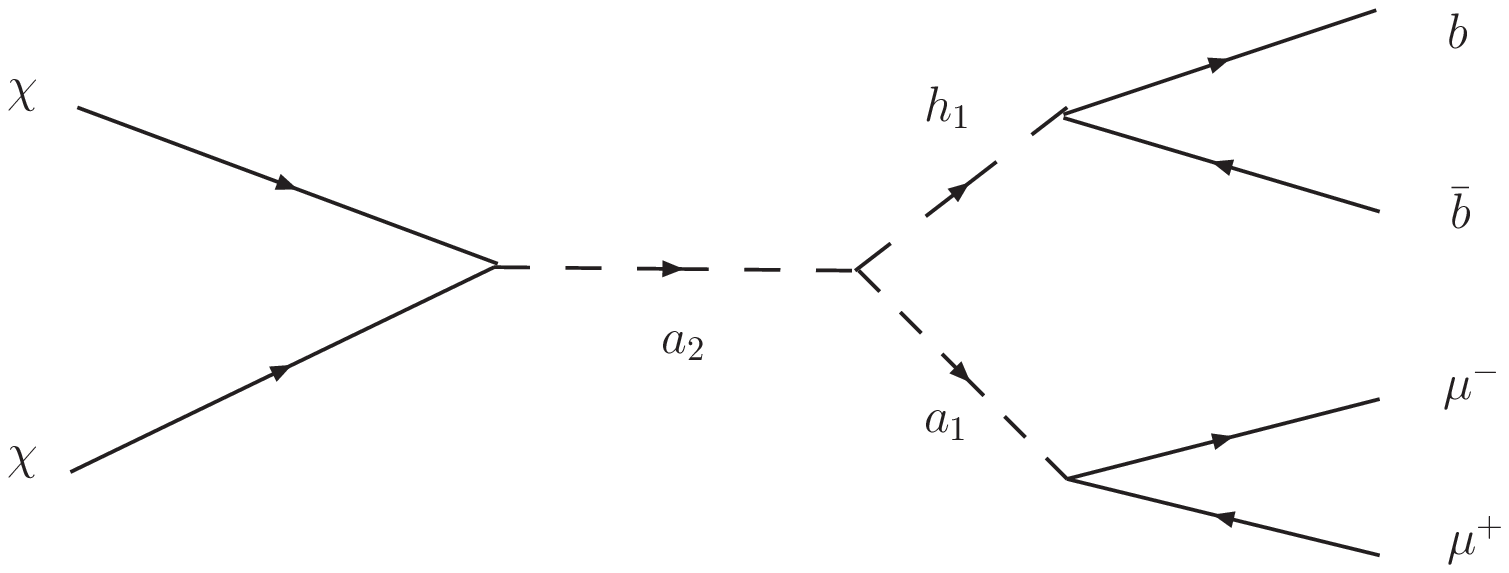} \\
\includegraphics[width=0.47\textwidth]{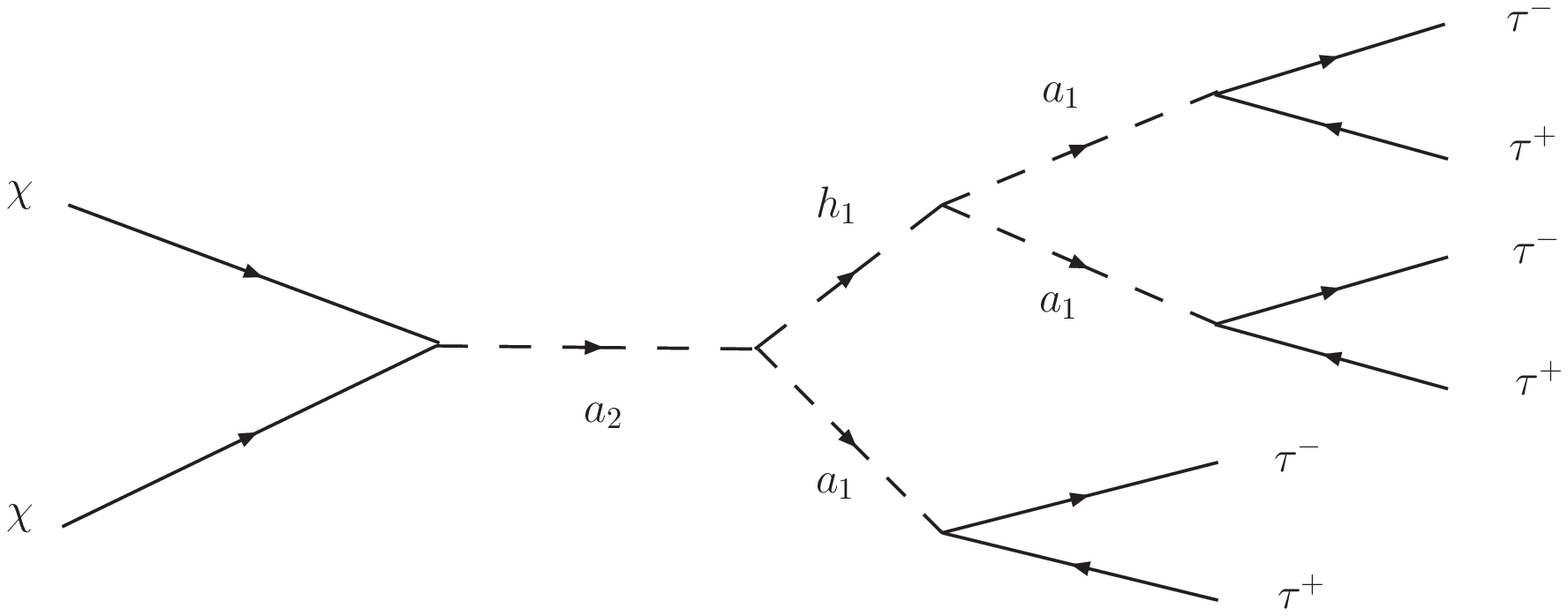} 
\caption{The Feynman diagram of the dominant neutralino annihilation channel. The upper panel is for the 
muon-favored case, while the lower panel is for the tau-favored case.}
\label{fig:annihi}
\end{figure}

Having established
$\chi\,\chi\rightarrow a_2 \rightarrow X$ as the main dark matter annihilation channel, 
the main annihilation products are equivalent to the decay products 
of $a_2$.
The main dark matter annihilation products are shown in Fig.~\ref{fig:annihi} for the two cases
exemplified by Table~\ref{tab:case1para} 
and Table~\ref{tab:case2para}.

To calculate the dark matter annihilation cross section, we will concentrate on the main annihilation 
channel $\chi\,\chi\,\rightarrow\, a_2\,\rightarrow a_1\,h_1$, which is true for both two cases. 
The annihilation rate is
\beqa
&&\sigma(\chi\,\chi\,\rightarrow\, a_2\,\rightarrow a_1\,h_1)\,v_{dm} \nonumber \\
&=&
\frac{g^2_{a_2\chi\chi}\,w^2_{a_1a_2h_1}}{16\,\pi\,\left[(M^2_{a_2}\,-\,4\,m^2_\chi)^2\,+\,\Gamma^2_{a_2}\,M^2_{a_2} \right]}\,\left(1\,-\,\frac{M^2_{h_1}}{4\,m^2_\chi}\right)\,, \nonumber \\
\label{eq:anni_cross_section}
\eeqa
up to ${\cal O}(v_{dm}^2/c^2)$ where the average dark matter speed is $v_{dm}/c\sim 10^{-3}$
in the galactic halo. Here the mass of $a_1$ is neglected and the 
coupling $w_{a_1a_2h_1}$ has mass dimension one. The co-annihilation effects can be neglected here, 
because other superpartner masses are much larger than the LSP mass. For the parameter points 
in Table~\ref{tab:case1para} and Table~\ref{tab:case2para}, we have $\sigma(\chi\,\chi\,\rightarrow a_1\,h_1)\,v_{dm}$ 
approximately 135~pb$\cdot$c and 301~pb$\cdot$c, respectively. Also from Table~\ref{tab:case1para} 
and Table~\ref{tab:case2para}, the width $\Gamma_{a_2}$ of $a_2$ is below 1 GeV, so the width part in the 
denominator of Eq.~(\ref{eq:anni_cross_section}) can be neglected for $|M_{a_2}\,-\,2\,m_\chi|\,>\,1$~GeV 
considered here. Therefore,  a large dark matter annihilation cross section can easily be obtained for the 
case of $m_\chi\,<\,m_t$. On the contrary, if $m_\chi\,>\,m_t$, the decay channel of $a_2\,\rightarrow\,t\,\bar{t}$ 
is open and the decay width of $a_2$ is of order 10~GeV. Then the dark matter annihilation cross section is limited 
by the width part in Eq.~(\ref{eq:anni_cross_section}), generically below 10~pb$\cdot$c and not large enough to 
explain the PAMELA data. Thus we have \textit{two} reasons to believe that the lightest neutralino mass is below the 
top quark mass: one is to have dominantly leptonic annihilation final states and the other one is to have a large annihilation cross section.

Our annihilation rates are by far larger than the necessary one ($\sim1$~pb$\cdot$c) to satisfy the dark matter 
thermal relic density. One possible explanation for this discrepancy is that the dark matter is nonthermal. For 
example, other long-lived particles like the gravitino or moduli can decay into the LSP at a later time~\cite{Moroi:1999zb}. 
As argued in~\cite{Moroi:1999zb}, a long-lived modulus field with a lifetime 
shorter than one second can naturally appear in the anomaly-mediated SUSY breaking model. The main dark matter 
relic abundance will be determined by the moduli annihilation cross section, which in principle can be smaller 
than the one for the neutralino and provide the observed dark matter relic density at the current time.

\section{Positron excesses from neutralino annihilation}
\label{sec:positrons}
We now calculate the positron 
spectrum out of dark matter annihilation and compare it with the PAMELA results to determine the PAMELA-favored 
parameter space in the NMSSM. We first discuss the source term of primary positrons from DM annihilations, 
propagation of cosmic ray positrons and then the positron fluxes measured at PAMELA. We also consider 
the charge-dependent solar modulation on the positron fraction spectrum.

\subsection{The source term for primary positrons}
\label{sec:primarypositrons}
For the two cases we are considering in this paper, the source term for primary positrons can be generally written as:
\beqa
q(\x, E)\,=\,\frac{1}{2}\,\langle \sigma\,v\rangle \left( \frac{\rho(\x)}{m_\chi}    \right)^2\,\frac{d N_{e^+}}{dE_{e^+}}\,. \label{eq:positronsource}
\eeqa
Here the overall factor $1/2$ is from the Majorana property of the neutralino DM candidate; $\langle \sigma\,v\rangle$ is the thermally 
averaged annihilation cross section and to a good  approximation can be replaced 
by the formula in Eq.~(\ref{eq:anni_cross_section}); $d N_{e^+}/dE_{e^+}$ 
is the energy spectrum of positrons; $\rho(\x)$ is the dark matter distribution inside the Milky Way halo. 
For the dark matter distribution we use either the Navarro, Frenk and White (NFW) profile~\cite{Navarro:1996gj} 
or the cored isothermal (ISO) profile~\cite{Bahcall:1980fb};
the use of other profiles may 
change discussions in this paper, especially for the gamma ray spectrum. 
The NFW and ISO profiles are parametrized as:
\beqa
\rho_{\rm NFW}(r)&=&\rho_{\odot}\,\left(\frac{r_\odot}{r}\right)\,\left( \frac{r_s\,+\,r_\odot}{r_s\,+\,r}     \right)^2\,,
\nonumber \\
\rho_{\rm ISO}(r)&=&\rho_{\odot}\,\left( \frac{r_s^2\,+\,r_\odot^2}{r_s^2\,+\,r^2}     \right)\,,
\eeqa
with $r_s\,=\,20$~kpc (NFW), 5 kpc (ISO) is the radius of the central core; 
$r_\odot\,=\,8.5$~kpc is the galactocentric distance of the solar system; 
$\rho_\odot\,=\,0.3$~GeV~cm$^{-3}$ is the solar neighborhood DM density. 

The positron energy spectrum function for the $\mu$ case can be generally expressed as
\begin{widetext}
\beqa
\frac{d N_{e^+}}{dE_{e^+}}&=&\int dE_{a_1}\,dE_{\mu^+}\,\frac{d N_{a_1}}{dE_{a_1}}\,\p(E_{a_1}\rightarrow E_{\mu^+})\,\p(E_{\mu^+}\rightarrow E_{e^+})   \nonumber \\
&+&2\, \int dE_{h_1}\,dE_{a_1}\,dE_{\mu^+}\,\frac{d N_{h_1}}{dE_{h_1}}\,\p(E_{h_1}\rightarrow E_{a_1})\,\p(E_{a_1}\rightarrow E_{\mu^+})\,\p(E_{\mu^+}\rightarrow E_{e^+})   \,.
\label{eq:dNdeformula}
\eeqa
\end{widetext}
Here $\p(E_i\rightarrow E_j)$ denotes the probability of a particle $i$ with energy $E_i$ decaying into a particle $j$ with energy $E_j$; 
the factor of 2 in the expression is because $h_1$ decays into 2 $a_1$'s. We neglect the positrons and electrons from the $b$ quark decays although 
the $h_1$ mainly decays into two $b$ quarks in the muon-favored case. 
Because positrons and electrons from $b$ decays are relatively soft, including them makes only a slight
change in the spectrum below 10~GeV, where the background is anyway dominant. 

From two $\chi$'s annihilating into $a_1$ and $h_1$, and neglecting the mass of $a_1$, we have
\beqa
\frac{d N_{a_1}}{dE_{a_1}}&=&{\rm Br}(a_2\,\rightarrow\,h_1\,a_1)\,\delta\left(E_{a_1}\,-\,(m_\chi\,-\,\frac{M^2_{h_1}}{4\,m_\chi})\right)\,,\nonumber \\
\frac{d N_{h_1}}{dE_{h_1}}&=&{\rm Br}(a_2\,\rightarrow\,h_1\,a_1)\,\delta\left(E_{h_1}\,-\,(m_\chi\,+\,\frac{M^2_{h_1}}{4\,m_\chi})\right)\,. \nonumber \\
\eeqa
Since the emission of muons in the $a_1$ rest frame and the emission of $a_1$'s in the $h_1$ rest frame  are isotropic, we have
\beqa
\p(E_{a_1}\rightarrow E_{\mu^+})&\approx&\frac{1}{E_{a_1}}\,\h(E_{a_1}\,-\,E_{\mu^+})\,,\nonumber \\
\p(E_{h_1}\rightarrow E_{a_1})&\approx&\frac{{\rm Br}(h_1\,\rightarrow\,a_1\,a_1)}{E_{h_1}}\,,
\eeqa
with $\h(x)$ is the heavy-side function and the masses of $\mu^+$ and $a_1$ are neglected in the 
approximation formula. Neglecting the muon polarization, the positron energy probability in muon decay
has the following analytic form~\cite{Cirelli:2008pk}
\beqa
\p(E_{\mu^+}\rightarrow E_{e^+}) &=&\frac{1}{3\,E_{\mu^+}}\left[5\,-\,9\,x^2\,+\,4\,x^3 \right]\, \nonumber \\
&& \times\,h(E_{\mu^+}\,-\,E_{e^+})\,,
\eeqa
with $x\equiv E_{e^+}/E_{\mu^+}$. Here the functions $\p(E_i\rightarrow E_j)$ are normalized such that $\int dE_j\, \p(E_i\rightarrow E_j) ={\rm Br}{( i \rightarrow j )}$. 

The $\tau$-favored case is more complicated than the $\mu$-case, 
because other than leptonic channels $\tau$ can also decay to 
various charged mesons, which decay eventually to electrons and positrons. 
In order to obtain the electron energy dependent probability from $\tau$ decays, 
we use {\tt PYTHIA}~ \cite{Sjostrand:2006za}, which calls the program TAUOLA~\cite{Jadach:1993hs}, 
to simulate the inclusive electron/positron energy spectrum from both direct and indirect $\tau$ decays.
\begin{figure}[ht!]
\includegraphics[width=0.45\textwidth]{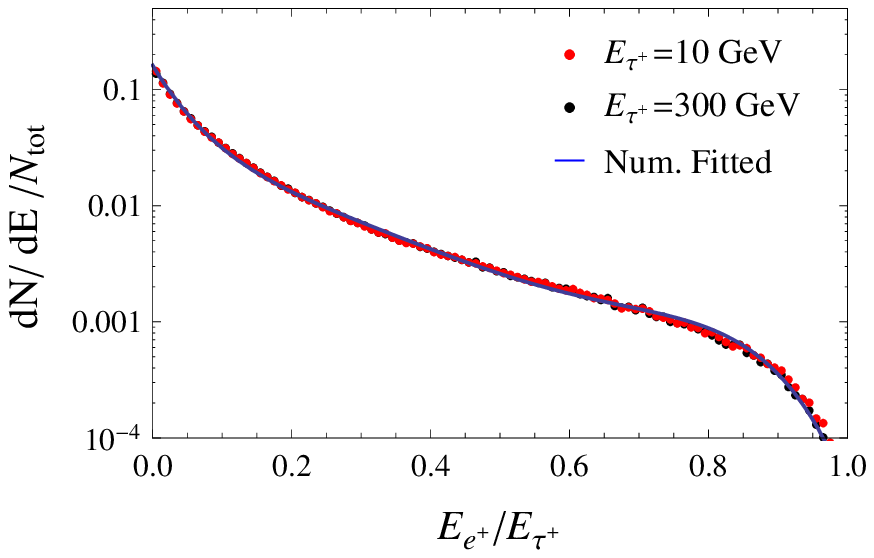} \hspace{3mm} \\
\includegraphics[width=0.45\textwidth]{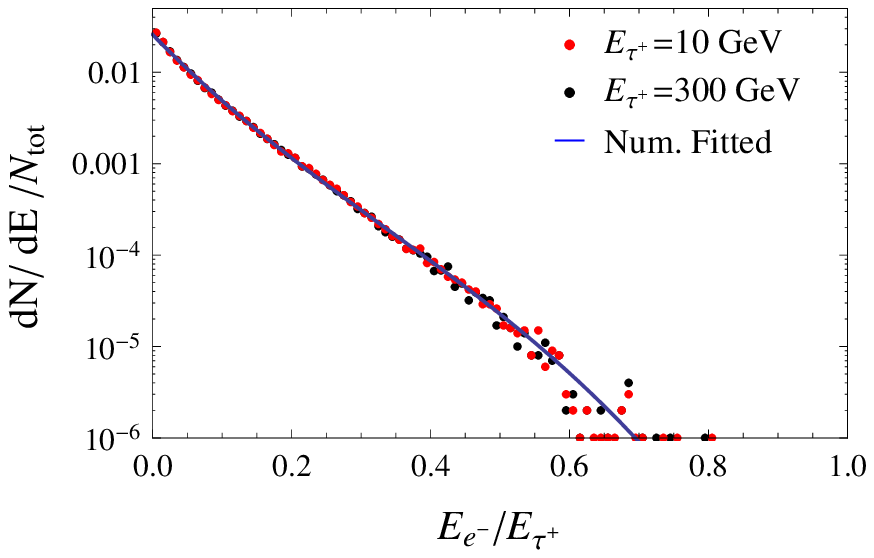}
\caption{Upper panel: the positron energy spectrum from inclusive tau decays. 
The red and the black points are for 10 GeV and 300 GeV $\tau^+$ energy, respectively. 
The blue line is from a fitted analytic function in the text. Lower panel: similar as 
the upper panel, but for electrons. 
On average, 1.16 positrons and 0.16 electrons are generated from one $\tau^+$ decay.}
\label{fig:taudecay}
\end{figure}
As can be seen from Fig.~\ref{fig:taudecay}, once the energy of the (unpolarized) $\tau$ 
is much larger than its mass, the fractional electron and positron energy spectra are independent 
of the energy of the $\tau$. The large fluctuations for $E_{e^-}/E_{\tau^+}>0.5$ in the 
lower panel of Fig.~\ref{fig:taudecay} are due to limited Monte Carlo statistics.
The tiny fraction of $e^-$ out of $\tau^+$ for $E_{e^-}/E_{\tau^+}>0.5$ 
can be understood from the fact that only high multiplicity final states can contain an $e^-$ with a different charge from $\tau^+$. 

For these $e^+$ and $e^-$ spectra from inclusive $\tau^+$ decays, the following fitted 
functions provide good agreement:
%
\beqa
&&\p(E_{\tau^+}\,\rightarrow\,E_{e^+})=\frac{1}{E_{\tau^+}}\,e^{-97.716\,x^5\,+\,223.389\,x^4} \nonumber \\
&\times&\,e^{-193.748\,x^3\,+\,82.595\,x^2\,-\,22.942\,x\,+\,2.783} \,\h(1\,-\,x)\,,\nonumber\\
&&\p(E_{\tau^+}\,\rightarrow\,E_{e^-})=\frac{1}{E_{\tau^+}}\,e^{-15.575\,x^3\,+\,15.79\,x^2}
\nonumber \\
&\times& e^{-\,18.083\,x\,+\,0.951}\,\h(1\,-\,x)\,,
\label{eq:taufragmentation}
\eeqa
%
with $x\,=\,E_{e^\pm}/E_{\tau^+}$. For each $\tau^+$, on average, 
there are $\int dE_{e^+} \p(E_{\tau^+}\,\rightarrow\,E_{e^+}) \approx 1.16$ 
positrons and $\int dE_{e^-} \p(E_{\tau^+}\,\rightarrow\,E_{e^-}) \approx 0.16$ 
electrons produced in the decay. When $a_1$ decays into $\tau^+\,+\,\tau^-$, 
there are on average 1.32 positrons out of a single $a_1$ decay. Therefore, 
we insert Eq.~(\ref{eq:taufragmentation}) into the analogous formula in Eq.~(\ref{eq:dNdeformula}) for the $\tau$ case. 
%
%

\subsection{Propagation of positrons and electrons}
\label{sec:propagation}
The propagation of positrons and electrons in the galactic medium is described by the following transport equation:
\beqa
\frac{\partial N}{\partial t}-\nabla\cdot[K(\x, E)\nabla N]-\frac{\partial}{\partial E} [b(E)N]=q(\x, E) \,, 
\eeqa
where $N(\x, E)$ denotes the positron number density per unit energy; $q(\x, E)$ 
is the positron source term; $K(\x, E)\,=\,K_0\,(E/E_0)^\delta$ is the diffusion 
constant with $E_0\,\equiv\,1$~GeV; $b(E)\,=\,E^2/(E_0\,\tau_E)$ is the positron 
energy synchrotron and inverse Compton loss rate with $\tau_E\,=\,10^{16}$~s. The 
diffusive halo is modeled as a cylinder with radius $r_s\,=\,20$~kpc and the 
vertical direction $z$ inside $(-L, L)$. The half thickness is not constrained and 
varies from 1 to 15~kpc. We will consider three different 
parameter points for the cosmic ray propagation model in Table~\ref{tab:propagationparameter}.
\begin{table}[htb]
\vspace*{0.3cm}
\renewcommand{\arraystretch}{1.6}
\begin{center}
\begin{tabular*}{0.5\textwidth} {@{\extracolsep{\fill}} lcccc} 
\hline \hline
Model & $\delta$  & $K_0$ [kpc$^2$/Myr] & $L$ [kpc]  \\ \hline 
M2     &     0.55    &     0.00595     &     1          \\   
MED   &     0.70     &     0.0112      &  4           \\
M1  &       0.46    &     0.0765     &   15       \\
\hline \hline
\end{tabular*}
\vspace{2mm}
\caption{Three combinations of cosmic ray propagation parameters, which give the 
minimum, median and maximum positron fluxes~\cite{Delahaye:2007fr}. }
\label{tab:propagationparameter}
\end{center}
\end{table}
Those sets of propagation parameters are compatible with the secondary/primary test for the secondary and primary antiprotons~\cite{Donato:2001ms}. 
Assuming a time-independent state and considering the diffusion constant as space independent, we have
\beqa
-\,K_0\,\left(\frac{E}{E_0}\right)^\delta\,\Delta\,N\, -\, \frac{\partial}{\partial E}\left[\frac{E^2}{E_0\,\tau_E}N\right]\,=\,q(\x, E) \,.
\label{eq:diffusion}
\eeqa
Defining a pseudo-time and a relative pseudo-time between the source point and the observation point, respectively, as
\beqa
&&\hat{t}(E)\,\equiv\,\tau_E\,\frac{(E/E_0)^{\delta\,-\,1}}{1\,-\,\delta}\,, \nonumber \\
&&
\hat{\tau}(E, E_S)\,=\,\hat{t}({E})\,-\,\hat{t}({E_S})\,,
\eeqa
the characteristic diffusion length in the radial direction is 
\beqa
\lambda_D(E, E_S)\,\equiv\,\sqrt{4\,K_0\,\hat{\tau}(E, E_S)}\,.
\eeqa
In Ref.~\cite{Delahaye:2007fr}, an analytic solution for the positron flux on the Earth has been obtained and has the following form:
\bea
\phi^\odot_{e^+}(E)&=&\frac{\beta\,c}{4\,\pi}\,\int^\infty_E\,d_{E_s}\,q(r_\odot, E_s)\,\times\,\frac{\tau_E\,E_0}{E^2} \nonumber \\
&&\,\times\,\eta\left(\lambda_D(E, E_s)\right)\,.
\eea
Using the Bessel  expansion method, the halo integral $\eta$, which is the volume integration of the Green function of Eq.~(\ref{eq:diffusion}), has both radial and vertical expansions: 
\begin{widetext}
\beqa
\eta(\lambda_D)\,=\,\sum_{i=1}^{\infty}\,\sum_{n=1}^{\infty}\,J_0(x_i\,r_\odot/r_s)\,\sin{(\frac{n\pi}{2})}\,{\rm exp}\left[-\left(\left(\frac{n\,\pi}{2\,L}\right)^2\,+\,\left(\frac{x_i}{r_s}\right)^2\right)\frac{\lambda_D^2}{4}\right] \,R_{i,n}\,,
\eeqa
with 
\beqa
R_{i, n}\,=\,\frac{2}{J_1(x_i)^2\,r_s^2}\,\int^{r_s}_0\,dr\,r\,J_0(x_n\,r/r_s)\,\frac{1}{L}\,\int^{+L}_{-L}\,dz\,\sin{(\frac{n\pi z}{2 L})}\,\left(\frac{\rho(\sqrt{r^2\,+\,z^2})}{\rho_{\odot}}\right)^2\,.
\eeqa
\end{widetext}
Here $J_k$ is the Bessel function of the first kind and $x_i$ is the $i$-th root of the $J_0$ Bessel function. In practice, one can use the numerically fitted functions in Ref.~\cite{Cirelli:2008id} to speed up the numerical calculations. For example, we use the following numerical function for the NFW dark matter profile and the M2 propagation model:
\beqa
\eta(\lambda_D)&=&0.5\,+\,0.774\,\tanh{\left( \frac{0.096\,-\,\ell}{0.211}   \right)}\, \nonumber \\
&\times& \left[  -0.448\, {\rm exp}\left( -\frac{(\ell\,-\,192.8)^2}{33.88}   \right)  \,+\,0.649           \right]\,,\nonumber \\
\eeqa
with $\ell\,\equiv\,\log_{10}(\lambda_D/{\rm kpc})$.

\subsection{Solar modulation}
\label{sec:solar}
PAMELA has measured the positron over electron fraction with energy below 10~GeV, and obtained a spectrum
significantly below the background fitted from other cosmic ray experiments. 
One possible explanation of this discrepancy is due to the charge sign dependence 
of the solar modulation. The magnetic field of the solar wind is dominated by the dipole term, 
and the projection of this dipole on the solar rotation axis can be either positive or negative, called $A^+$ and $A^-$ states, respectively. At each sunspot maximum, the dipole reverses its direction 
and leads to a periodic function for the dipole magnetic field with a roughly 
12 year period.~\footnote{Figure 1 of Ref.~\cite{Clem} shows clear evidence for this behavior in the electron flux.}

Using two functions $c_+(E)$ and $c_-(E)$ to model the solar modulation, we have the observed positron fraction on the Earth as:
\beqa
F^{\pm}_{\oplus}(E)\,=\,\frac{c_{\pm}(E)\,\Phi_{e^+}(E)}{c_{\pm}(E)\,\Phi_{e^+}(E)\,+\,c_{\mp}(E)\,\Phi_{e^-}(E)}\,,
\eeqa
with $+$ for the solar system in the $A^+$ state and $-$ for the $A^-$ state. Here $\Phi_{e^\pm}(E)$ denote the positron/electron fluxes. One notices that only the ratio of $c_+(E)/c_-(E)$  is relevant for the positron fraction. 
The ratio of the total electron flux in the $A^+$ cycle to total electron flux at a similar phase in the $A^-$ cycle is 
\beqa
R(E)\,=\,\frac{c_{+}(E)\,\Phi_{e^+}(E)\,+\,c_{-}(E)\,\Phi_{e^-}(E)}{c_{-}(E)\,\Phi_{e^+}(E)\,+\,c_{+}(E)\,\Phi_{e^-}(E)}\,.
\eeqa
This ratio can be fitted in principle 
from the data shown in Fig. 5 of Ref.~\cite{Clem}. 
However, since the lower energy bins of PAMELA (where the contribution from dark matter is negligible) 
are measured so precisely, one can use the data from PAMELA ($A^-$ cycle) and the known background~\cite{Moskalenko:1997gh} 
(without an assumption of solar modulation) to determine this ratio. The following fitted formula can achieve the goal:
\beqa
R(E)={\rm max}\left[{\rm min}\left(0.48+0.2\log(E/{\rm GeV}), 1.0\right), 0.2\right]\,,
\eeqa
which will be used in our following analysis about the comparison of model predictions and PAMELA data. 
The general form of this expression is suggested by Ref.~\cite{Baltz:1998xv}. 
Eventually one needs to understand or to calculate this ratio function or $c_\pm(E)$ from first principles. 
Due to the magnitude of the solar magnetic field, only positrons/electrons with energy below about 10 GeV can 
be influenced by the solar wind. $R(E)$ is a monotonically increasing function and saturates at unity above $7$~GeV.

In terms of $R(E)$ and $F(E)\,\equiv\,\Phi_{e^+}(E)/(\Phi_{e^-}(E)\,+\,\Phi_{e^+}(E))$, which is the positron fraction without solar modulation effects, one has:
\beqa
F^-_\oplus(E)&=&\frac{F^2\,(R\,+\,1)\,-\,F\,R}{2\,F\,-\,1}\,,\nonumber \\
 F^+_\oplus(E)&=&\frac{F^2\,(R\,+\,1)\,-\,F}{R\,(2\,F\,-\,1)}\,.
\eeqa
%

\subsection{PAMELA positron excess}
\label{sec:pamela}
Before we compare the positrons from the neutralino annihilations in the 
NMSSM to the PAMELA results, we first discuss the electron and positron 
backgrounds from standard astrophysicial processes. Background positrons are mostly secondaries 
originating from spallation processes of cosmic rays, mostly primary protons, 
off the interstellar gas, thus mainly occurring in the galactic disk. 
The primary electrons are mainly produced by shock wave acceleration 
in supernovae. To simplify our comparison of model predictions to experimental 
data (without the solar modulation effects), we use the following numerically 
fitted formulae for the background~\cite{Baltz:1998xv}, which agrees with the 
full results calculated in~\cite{Moskalenko:1997gh}:
\beqa
\phi^{e^-}_{\rm prim, bkg}&=&\frac{0.16\,E^{-1.1}}{1\,+\,11\,E^{0.9}\,+\,3.2\,E^{2.15}}\,, \nonumber
\\
\phi^{e^-}_{\rm sec, bkg}&=&\frac{0.70\,E^{0.7}}{1\,+\,110\,E^{1.5}\,+\,600\,E^{2.9}\,+\,580\,E^{4.2}}\,,  \nonumber
\\
\phi^{e^+}_{\rm sec, bkg}&=&\frac{4.5\,E^{0.7}}{1\,+\,650\,E^{2.3}\,+\,1500\,E^{4.2}}\,, 
\label{eq:electronBG}
\eeqa
in ${\rm GeV}^{-1}\,{\rm cm}^{-2}\,{\rm s}^{-1}\,{\rm sr}^{-1}$ and with $E$ in GeV. Combining the background and signal positrons, we have a general formula for the positron fraction 
\begin{widetext}
\beqa
F^-_\oplus\,=\,\frac{c_-\,\phi(e^+)}{c_+\,\phi(e^-)\,+\,c_-\,\phi(e^+)}
\,=\,\frac{c_-\,(\phi^{e^+}_{\rm sec, bkg}\,+\,\phi^{e^+}_{\rm sig})}{c_+(\phi^{e^-}_{\rm prim, bkg}+\phi^{e^-}_{\rm sec, bkg}+\phi^{e^-}_{\rm sig})+c_-(\phi^{e^+}_{\rm sec, bkg}+\phi^{e^+}_{\rm sig})}\,,
\eeqa
\end{widetext}
assuming that the PAMELA data were taken when the sun is in the $A^-$ cycle.
 
\begin{figure}[ht]
\centerline{ \hspace*{-1cm}
\includegraphics[width=0.48\textwidth]{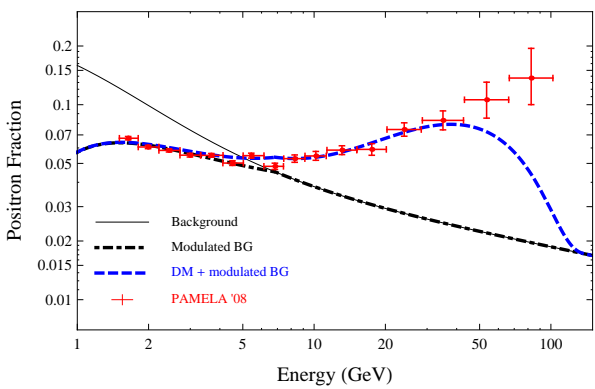}
} 
\caption{The positron fraction from the $\mu$-favored model point in the NMSSM. 
The solid black line is the background without considering solar modulation. 
The dotdashed black line is the background with the solar modulation effects. 
The dashed blue line is the positron fraction from neutralino ($m_\chi=162$ GeV) 
annihilations plus the modulated background. The red points are data from PAMELA 
with one standard deviation errors. The dark matter annihilation cross section  
is $6.0 \times 10^{-24}$ cm$^3$ s$^{-1}$. The M2 propagation model is used here.}
\label{fig:muon}
\end{figure}

Using the model point and the masses of $\chi$, $h_1$ and $a_1$ reported in Table~\ref{tab:case1para}, 
we have the positron excess for the $\mu$-favored model point in the NMSSM shown in Fig.~\ref{fig:muon}. To generate the plot in Fig.~\ref{fig:muon}, we have used the M2 propagation model 
from Table~\ref{tab:propagationparameter}, which provides a best fit to the PAMELA data. 
The other two propagation models generate a flatter curve than the M2 model. This is 
because as the thickness of the diffusive halo decreases, the positrons detected 
on the Earth originate from a nearby region (the characteristic propagation distance 
$\lambda_D$ decreases), and hence low-energy positrons are less likely to reach the 
Earth. This leads to a steeper spectrum for the M2 propagation model. Since all 
three propagation models are supported by the N-body simulation, the combination of 
the NMSSM and the PAMELA data (assuming the NMSSM interpretation was confirmed by, e.g.
collider discoveries) could help to determine a correct galaxy model. 

In Fig.~\ref{fig:muon} we note that the NMSSM model is 3 sigma below the
PAMELA data point in the highest energy bin. However the overall agreement
is quite good. To estimate the goodness of fit conservatively, we ignore
the PAMELA data below 7.4 GeV, which should have negligible 
contribution from the dark matter annihilation, and
calculate the averaged $\chi^2$ for the 8 bins above 7.4 GeV:
\beqa
\frac{\chi^2}{8}\,=\,\frac{1}{8}\,\sum_{i\,=\,1}^{8}\,\left(\frac{x^{\rm model}_i\,-\,x_i^{\rm exe}}{\sigma_i}\right)^2\,\approx\,0.9\,.
\eeqa

If the neutralino annihilation in the NMSSM is the explanation 
for the PAMELA data, the rising feature of the positron fraction spectrum should end at around 70 GeV. This is a dramatic prediction for future PAMELA results. This predicted turnover follows uniquely from the requirement in the NMSSM model that the dark matter LSP mass cannot exceed the top quark mass.

\begin{figure}[ht!]
\centerline{ \hspace*{0cm}
\includegraphics[width=0.45\textwidth]{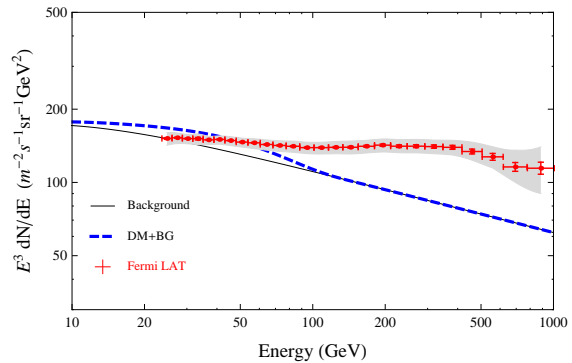} 
} 
\caption{The electron plus positron spectrum from the $\mu$-favored model.  
The solid black line is the background. The dashed blue line is from 
the neutralino annihilation plus the background.  
The red crossed points are the results from Fermi Large Area Telescope  
(Fermi LAT) with the gray band for systematic errors. 
The dark matter annihilation cross section is $6.0\times 10^{-24}$ cm$^2$ s$^{-1}$. 
The M2 propagation model is used here.}
\label{fig:muonelectron}
\end{figure}

For completeness, we also show the electron plus positron energy spectrum 
from dark matter annihilation in Fig.~\ref{fig:muonelectron}.
We also include the latest results from Fermi LAT 
in the red and crossed points~\cite{FermiLAT} in Fig.~\ref{fig:muonelectron}.  
There is an additional uncertainty from the LAT energy scale, which can shift 
the whole gray band by 5\% (up) to 10\% (down) and is not shown in this figure. 
In this figure the electron and positron fluxes from the background in Eq.~(\ref{eq:electronBG}) plus signal have been normalized to agree with the first bin of Fermi LAT. The positron fraction predictions from PAMELA  are unchanged by this manipulation. 

The agreement between the predicted electron+positron spectrum and the Fermi LAT data is about the same with and without adding the NMSSM signal. The
generally poor agreement should thus be attributed to a defect in our
understanding of the cosmic electron/positron background. 
More generally, we conclude that a neutralino annihilation explanation for 
PAMELA is consistent with the Fermi LAT results, as long as the extra
contribution to the electron+positron spectrum is within the Fermi LAT errors.

\begin{figure}[ht!]
\includegraphics[width=0.45\textwidth]{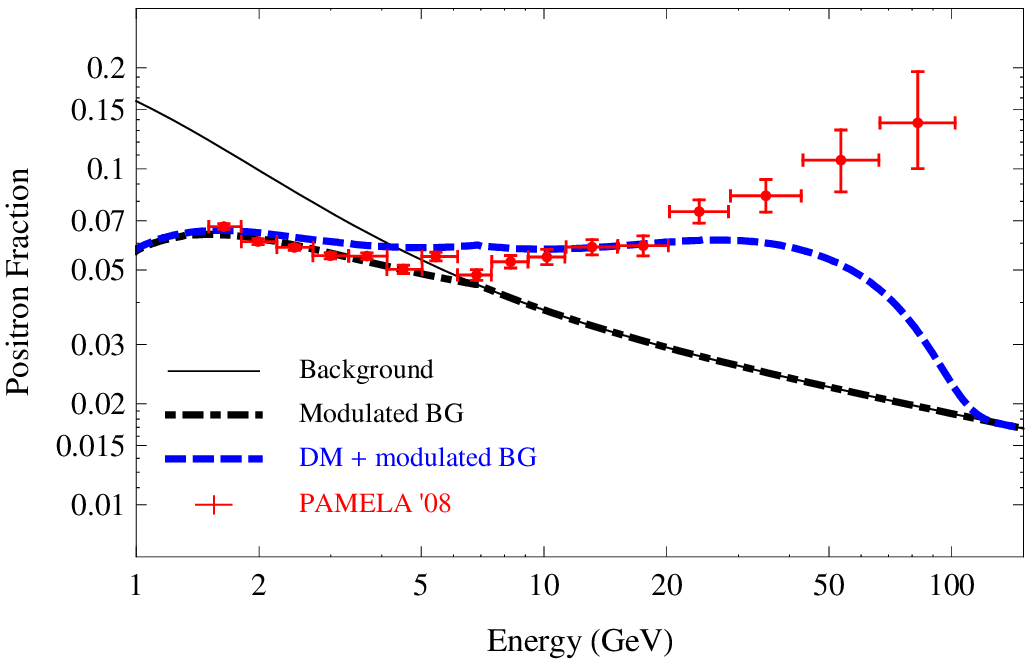} \\
\includegraphics[width=0.43\textwidth]{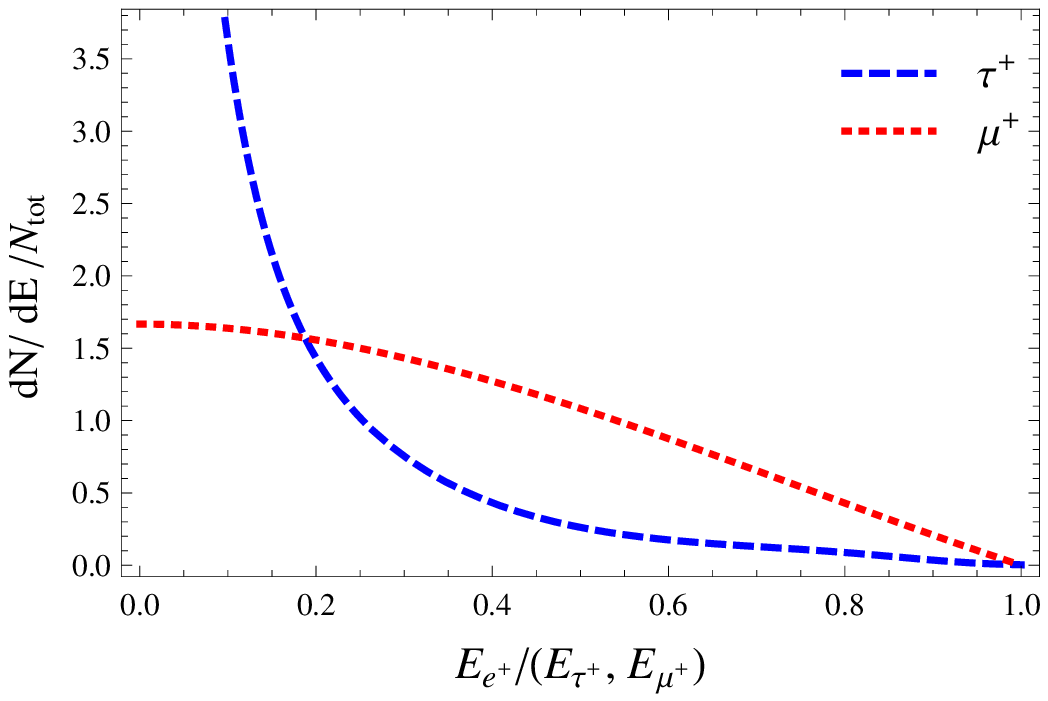}
\caption{Upper panel: the positron fraction from one $\tau$-favored model point in the NMSSM. 
The dark matter ($m_\chi=160$~GeV) annihilation cross section is $9.0\times 10^{-24}$ cm$^2$ s$^{-1}$. 
The M2 propagation model is used here. Lower panel: a comparison of positron energy 
spectra from $\tau^+$ decays and from $\mu^+$ decays.}
\label{fig:tau}
\end{figure}

Having discussed the $\mu$-favored point in the NMSSM, we also report the results for the $\tau$-favored point. 
It turns out that although $\tau$-favored points can be found easily in the parameter space of the NMSSM, 
they provide a worse fit to the PAMELA data. This is mainly because the positrons from $\tau^+$ decays 
are softer than the positrons from $\mu^+$ decays. This fact can be seen from the lower panel of Fig.~\ref{fig:tau}, 
where a comparison between $\mu$ and $\tau$ cases is shown. 
Only taking the PAMELA data above $7.4$~GeV into account, we calculate the average $\chi^2$ for 
the fit of this $\tau$-favored model point to PAMELA data as $\chi^2/8\,\approx\,3.4$. Because 
of this large $\chi^2$, we will 
concentrate on the $\mu$-favored model from here on.

\section{PAMELA antiproton}
\label{sec:pamela-proton}
\begin{figure}[ht!]
\centerline{ \hspace*{0cm}
\includegraphics[width=0.46\textwidth]{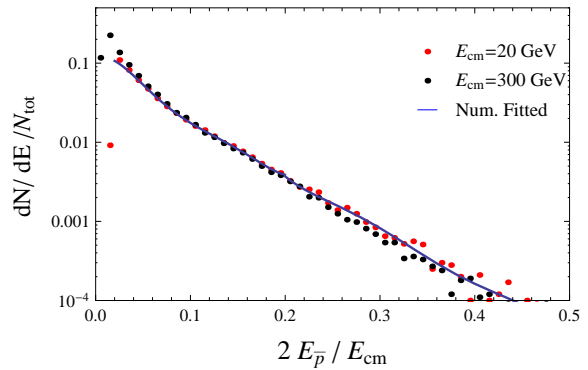}
} 
\caption{The  $b\,\bar{b}$ to antiproton fragmentation function.
The red and black points are for 20~GeV and 300~GeV $b\,\bar{b}$ center of mass energies. The blue line is from a 
fitted analytic function described in the text. The fragmentation function is extracted using {\tt PYTHIA}.}
\label{fig:bbarfrag}
\end{figure}

Since there is no excess of the antiproton fraction observed at PAMELA, this imposes constraints 
on the antiproton production cross section from dark matter annihilation. Specifically to the 
muon-favored case in the NMSSM, the dominant source of hadronic production is from  $h_1\,\rightarrow\,b\,\bar{b}$ 
and $a_2\,\rightarrow\,b\,\bar{b}$. Similarly to the calculations for the positron fraction spectrum, we first use an analytic function to fit the 
fragmentation function of $b\,\bar{b}$ to antiprotons. The antiproton energy spectra 
for two different $b\,\bar{b}$ center of mass energies are extracted using {\tt PYTHIA} and shown in Fig.~\ref{fig:bbarfrag}.
We use the following fitted function later to replace the numerically simulated antiproton energy spectrum from $h_1$ decays:
\beqa
&&\p(E_{h_1}\,\rightarrow\,b\bar{b}\,\rightarrow\,T_{\bar{p}})=\frac{2}{E_{h_1}} \,e^{-307.05\,x^5+525.86\,x^4}\nonumber \\
&&\hspace{1cm}\times\,e^{-331.51\,x^3+96.71\,x^2-28.56\,x+2.755} \,,
\eeqa
where $x\,=\,2\,(T_{\bar{p}}+m_p)/E_{h_1}$. As is customary in antiproton cosmic ray analyses, 
we substitute the proton energy by its kinetic energy $T_{\bar{p}}\,=\,E_{\bar{p}}\,-\,m_p$ in the following.  
The antiproton spectrum from $a_2$ has a same formula replacing $E_{h_1}$ with $E_{a_2}$.

The source term for antiprotons from DM annihilations is similar to Eq.~(\ref{eq:positronsource}) but substituting $dN_{e^+}/dE_{e^+}$ with
\beqa
&&\frac{d N_{\bar{p}}}{dT_{\bar{p}}}=\int dE_{h_1}\,\frac{d N_{h_1}}{dE_{h_1}}\,{\rm Br}(h_1\,\rightarrow\,b\bar{b}) \nonumber \\
&&\times\p(E_{h_1}\rightarrow b\bar{b} \rightarrow T_{\bar{p}})+\int dE_{a_2}\,\frac{d N_{a_2}}{dE_{a_2}}\,{\rm Br}(a_2\,\rightarrow\,b\bar{b})\nonumber \\
&&\times \p(E_{a_2} \rightarrow b\bar{b} \rightarrow T_{\bar{p}})\,.
\eeqa
However, due to the fact that $m_p\,\gg\,m_e$, the energy loss term for antiprotons can be neglected. 
The steady diffusion equation for antiprotons is~\cite{Donato:2001ms}:
\beqa 
-\,K_p(T)\,\Delta\,N &+&\frac{\partial}{\partial z}({\rm sign}(z)\,V_c\,N) \nonumber \\
&+&\,2\,h\,\delta{(z)}\,\Gamma_{\rm ann}\,N \,=\,q(\x, T) \,.
\label{eq:antiprotondiffusion}
\eeqa
Here $N$ is the number density of antiprotons per unit energy and $K_p(T)\,=\,K_0\,\beta\,(p/{\rm GeV})^\delta$ 
with $\beta$ and $p$ are the antiproton velocity and momentum. The second term is related to the convective wind, 
which has a direction outward from the galactic plane and represents the movement of the medium responsible for the antiproton diffusion.
\begin{table}[htb]
\vspace*{0.3cm}
\renewcommand{\arraystretch}{1.6}
\begin{center}
\begin{tabular*}{0.5\textwidth} {@{\extracolsep{\fill}} ccccc} 
\hline \hline
Model & $\delta$  & $K_0$ [kpc$^2$/Myr] & $L$ [kpc]  & $V_c$ [km/s]  \\ \hline 
M2    &    0.55     &    0.00595     &     1       &  13.5   \\   
MED   &     0.70     &     0.0112      &  4         &   12   \\
MAX   &       0.46    &     0.0765     &   15      &  5  \\
\hline \hline
\end{tabular*}
\vspace{2mm}
\caption{Three combinations of cosmic ray propagation parameters, which give the minimum, median and maximum signal antiproton fluxes. }
\label{tab:propagationparameter2}
\end{center}
\end{table}
The velocity $V_c$ is assumed to be constant and has values shown in Table~\ref{tab:propagationparameter2} 
for different propagation models. The third term represents annihilations of antiprotons and interstellar 
protons in the galactic plane with a thickness $h\,=\,0.1$~kpc. The annihilation rate between antiproton and protons  
is $\Gamma_{\rm ann}\,=\,(n_{\rm H}\,+\,4^{2/3}\,n_{\rm He})\,\sigma^{\rm ann}_{p\bar{p}}\,v_{\bar{p}}$ 
with $n_{\rm H}\,\approx\,1$~cm$^{-3}$ and $n_{\rm He}\,\approx\,0.07$~cm$^{-3}$. The $\sigma^{\rm ann}_{p\bar{p}}$ as 
a function of antiproton kinetic energy is given in~\cite{Tan:1983de} 
and \cite{Protheroe:1981gj}: $661(1\,+\,0.0115\, T^{-0.774}\,-\,0.984\, T^{0.0151})$~mb 
for $T\,<\,15.5$~GeV and $36 \,T^{-0.5}$~mb for $T\,\ge \,15.5 $~GeV. This annihilation 
process is dominant at low energy and leads to a decreased flux of antiprotons with low energy. 
In our analysis, other non-annihilation interactions between antiprotons and the interstellar medium 
are neglected. These effects are not important for the antiproton flux with energy above a few GeV.

Similar to the positron case, the diffusion equation for the antiproton can be solved analytically and 
has the following concise form as its solution~\cite{Donato:2001ms}:
\beqa
\phi^\odot_{\bar{p}}(T_{\bar{p}})\,=\,\frac{\beta_{\bar{p}}\,c}{4\,\pi}\,\left(\frac{\rho_\odot}{m_\chi}\right)^2\,\frac{1}{2}\,\langle \sigma\,v\rangle\,\frac{d N_{\bar{p}}}{dT_{\bar{p}}}\,\times\,R(T_{\bar{p}})\,.
\eeqa
For the M2 propagation model and the NFW dark matter profile, we numerically fit an analytical formula for $R(T_{\bar{p}})$
\beqa
  R(T)/{\rm Myr}&=&10^{1.352\,+\,0.0542\,\log_{10}(T)\,-\,0.265\,\log_{10}(T)^2} \nonumber \\
&\times& 10^{\,+\,0.0597\,\log_{10}(T)^3\,-,0.00575\,\log_{10}(T)^4}\,, \nonumber \\ 
\eeqa
with $T$ in GeV. To obtain the antiproton flux observed on the Earth, we need to take the solar modulation effect into account. 
The solar modulation effect is represented by a parameter $\phi$, which is 500~MV for minimum 
solar activity when PAMELA was taking data. The energy spectrum of antiprotons on the Earth is
\beqa
\phi^\oplus_{\bar{p}}(T_{\bar{p}})&=&\phi^\odot_{\bar{p}}(T_{\bar{p}}\,+\,|Z|\,\phi) \nonumber \\
&\times& \frac{(T_{\bar{p}}\,+\,m_p)^2\,-\,m_p^2}{(T_{\bar{p}}\,+\,|Z|\,\phi\,+\,m_p)^2\,-\,m_p^2}\,,
\eeqa
with $Z=1$ the electric charge of the antiproton. 

The background for primary protons can be extrapolated from other cosmic ray experiments. Using the data from AMS~\cite{Alcaraz:2000vp},  we arrive at the following fitted analytic function to describe the primary proton background:
\beqa
&&\phi^{\rm BG, AMS}_p(T)=e^{-0.00097\,\log^5{T}\,+\,0.012\,\log^4{T}} \nonumber \\
&\times& e^{\,+\,0.014\,\log^3{T}\,-\,0.382\,\log^2{T}\,-\,0.828\,\log{T}\,+\,6.88}\,,
\eeqa
with $T$ in GeV and the flux in m$^{-2}$\,s$^{-1}$\,sr$^{-1}$\,GeV$^{-1}$. Similarly, from  CAPRICE98~\cite{Boezio:2002ha} data, we  have
\beqa
&&\phi^{\rm BG, CAPRICE}_p(T)=e^{-0.0005\,\log^5{T}\,+\,0.005\,\log^4{T} }\nonumber \\
&\times& e^{+0.019\,\log^3{T}\,-\,0.433\,\log^2{T}\,-\,0.882\,\log{T}\,+\,6.89}\,,
\eeqa
The primary proton fluxes measured at AMS and CAPRICE can deviate from each other 
by a difference as large as 20\%. We include both results  in our analysis to encode 
uncertainties of our primary proton flux  background. The secondary antiproton background 
can be found in the detailed analysis in~\cite{Bringmann:2006im} and fitted in~\cite{Cirelli:2008id} as:
\beqa
\phi^{\rm BG}_{\bar{p}}(T)&=&10^{0.028\,\log_{10}^4{T}\,-\,0.02\,\log_{10}^3{T}}
\nonumber \\
&\times&10^{\,-\,1.0\,\log_{10}^2{T}\,+\,0.07\,\log_{10}{T}\,-\,1.64}\,.
\eeqa
\begin{figure}[ht!]
\centerline{ \hspace*{0cm}
\includegraphics[width=0.50\textwidth]{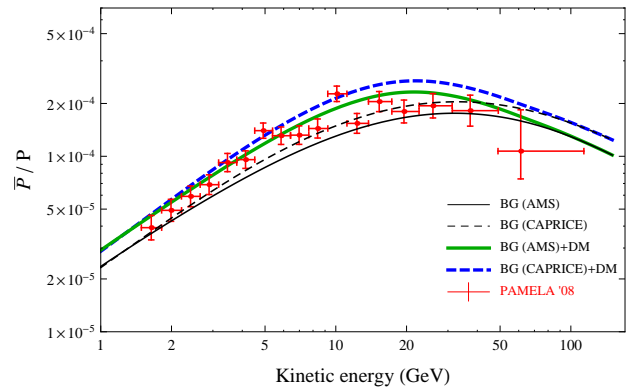}
} 
\caption{The antiproton-to-proton flux ratio as a function of kinetic energy. 
The black solid and dashed lines are the backgrounds from AMS and CAPRICE, respectively.  
The thick green solid line is from the dark matter annihilation plus the background (AMS) 
using the M2 propagation model. The thick blue dashed line is using the background (CAPRICE). 
The red crossed points are data from PAMELA. The neutalino mass is $m_\chi\,=\,162$~GeV and 
its annihilation cross section is  $6.0 \times 10^{-24}$ cm$^2$ s$^{-1}$. }
\label{fig:bbar}
\end{figure}
A comparison of the antiproton/proton flux ratio from the neutralino annihilation and the 
PAMELA data is shown in Fig.~\ref{fig:bbar}.  Noticing that the M2 propagation model provides 
a better fit to the PAMELA positron data, the model point in Table~\ref{tab:case1para} is 
marginally allowed by the PAMELA antiproton data. To quantify the discrepancy between the model 
prediction and the PAMELA observed data, we calculate the  $\chi^2$ by including all bins.  
We also calculate the $\chi^2$ between the background and PAMELA data. Using the background 
extrapolated from AMS, we have $\chi^2$ to be 1.6 for the DM prediction plus the background  
and 3.6 for the background only. While, using the background from CAPRICE, we have $\chi^2$ to be 4.0 for the DM prediction plus the background  and 2.2 for the background only. We conclude that the 
model point in Table~\ref{tab:case1para} is allowed by the PAMELA antiproton data, taking into
account the large uncertainties of the background primary proton flux. 

Although the branching ratio ${\rm Br}(a_2\,\rightarrow\,b\,\bar{b})$ 
is smaller than ${\rm Br}(a_2\rightarrow h_1\,a_1){\rm Br}(h_1\rightarrow b\,\bar{b})$, 
the center of mass energy of $b\,\bar{b}$ directly out of $a_2$ is approximately twice 
of the center of mass energy out of $h_1$. Therefore, the large antiproton fraction 
for the kinetic energy above 10~GeV mainly comes from dark matter annihilating directly
into $b\,\bar{b}$. To suppress the antiproton flux more efficiently, one could change the 
model parameters to suppress the branching ratio of $a_2$ to $b\,\bar{b}$. One simple way to do this is
to reduce $\tan{\beta}$. However, by doing so, the branching ratio of $h_1\,\rightarrow a_1\,a_1$ is 
also increased, creating a tension with the upper limit from D0 described in Section~\ref{sec:TEVATRON}.

\section{Gamma ray fluxes for Fermi LAT}
\label{sec:gammaray}
If the $\mu$-favored model points in the NMSSM are the correct explanation of  the PAMELA positron excess, there will be lots of associated gamma rays generated. Existing gamma ray data from HESS~\cite{HESS} and EGRET~\cite{EGRET} can in principle impose constraints on the dark matter annihilation cross section to electrons and positrons. Also, the recent and upcoming gamma-ray flux data from Fermi LAT have a smaller statistic uncertainty and can be used to test the PAMELA-favored NMSSM model.  

The differential gamma-ray flux from the dark matter annihilation has the following general formula:
\beqa
\frac{d^2\Phi_\gamma}{d\Omega\,dE_\gamma}\,=\,\frac{1}{2}\,\frac{\langle \sigma v \rangle}{4\,\pi\,m_\chi^2}\,\frac{dN_\gamma}{dE_\gamma}\,\int^\infty_0\,\rho^2(r)\,dl(\psi)\,.
\eeqa
Here $r^2\,=\,l^2+r^2_\odot-2\,l\,r_\odot\,\cos{\psi}$ with $\psi$ as 
the angle between the line of sight and the galactic plane. One can 
separate the astrophysical uncertainties by introducing the quantity 
\beqa
J(\psi)\,=\,\frac{1}{r_\odot\,\rho^2_\odot}\,\int^\infty_0\,\rho^2(r)\,dl(\psi)\,.
\eeqa
Performing the solid angle integration, the differential gamma-ray  flux is
\beqa
\frac{d\Phi_\gamma}{dE_\gamma}\,=\,\frac{1}{2}\,\frac{r_\odot\,\rho^2_\odot\,\langle \sigma v \rangle}{4\,\pi\,m_\chi^2}\,\frac{dN_\gamma}{dE_\gamma}\,\bar{J}(\Delta \Omega)\,\Delta \Omega\,,
\eeqa
with $\bar{J}(\Delta \Omega)\,\equiv (1/\Delta \Omega)\,\int_{\Delta \Omega} J(\psi)\,d\Omega$ and $\Delta \Omega=2\pi(1-\cos{\psi})$ for the region around the galactic center. For example, we have $\bar{J}(\Delta \Omega)\,\Delta \Omega \approx 1$ for $\Delta \Omega=10^{-3}$ and $\bar{J}(\Delta \Omega)\,\Delta \Omega \approx 0.1$ for $\Delta \Omega=10^{-5}$~sr using the NFW dark matter profile~\cite{Ponton:2008zv}. 

The gamma-rays from the $\mu$-favored model point of the NMSSM have two sources: one is related to the muons in the final state and the other one is related to the bottom quarks in the annihilation final state. Altogether, we have 
\beqa
\frac{dN_\gamma}{dE_\gamma}&=&{\rm Br}(a_2\,\rightarrow\,a_1\,h_1)\frac{dN^\mu_\gamma(E_{a_1})}{dE_\gamma} \nonumber \\
&+&{\rm Br}(a_2\,\rightarrow\,a_1\,h_1)\frac{dN^b_\gamma(E_{h_1})}{dE_\gamma} 
\nonumber \\
&+& {\rm Br}(a_2\,\rightarrow\,b\,\bar{b})\frac{dN^b_\gamma(E_{a_2})}{dE_\gamma}\,.
\eeqa
There are two processes to generate gamma-rays associated with the muon final state. One is through final state radiation (FSR) and the other one is from the radiative muon decays into photons. For the final state radiation, we have 
\beqa
&&\frac{dN^\mu_{\gamma, FSR}(E_{a_1})}{dE_\gamma}\,=\,\frac{2}{E_{a_1}}\,\int^{\frac{M_{a_1}-2m_\mu}{M_{a_1}-m_\mu}}_{\frac{E_\gamma}{E_{a_1}}}\,dx\,\frac{1}{x}\,
\frac{\alpha}{\pi} \nonumber \\
&\times& \left(\frac{1\,+\,(1\,-\,x)^2}{x}     \right)\left( \log\left(\frac{M^2_{a_1}(1-x)}{m_\mu^2}\right)\,-\,1    \right)\,. \nonumber \\
\eeqa
For the radiative muon decays: $\mu^-\rightarrow e^-\,\nu_\mu\,\bar{\nu}_e\,\gamma$ and $\mu^+\rightarrow e^+\,\nu_e\,\bar{\nu}_\mu\,\gamma$, one has~\cite{Essig:2009jx}
\beqa
&&\frac{dN^\mu_{\gamma, RAD}(E_{a_1})}{dE_\gamma}\,=\, \nonumber \\
&&\quad\frac{2}{E_{a_1}}\,\int^{1}_{\frac{E_\gamma}{E_{a_1}}}\,dx\,\frac{2}{x}\,\int^{{\rm min}(1, \frac{2\,x}{1-\beta})}_{\frac{2\,x}{1+\beta}} dy\,\frac{1}{y}\,{\cal F}(y)\,.
\eeqa
with  $\beta\,=\,\sqrt{1-4m^2_\mu/m^2_\phi}$ and ${\cal F}(y)$ as the photon spectrum in the muon rest frame and given by
\beqa
{\cal F}(y)&=&\frac{\alpha}{3\,\pi}\,\frac{1\,-\,y}{y}\left( (3\,-\,2\,y\,+\,4\,y^2\,-\,2\,y^3)\log{\frac{m_\mu^2}{m_e^2}} \right.
\nonumber \\
&&\left. \,-\,\frac{17}{2}\,+\,\frac{23\,y}{6}\,-\,\frac{101\,y^2}{12}\,+\,\frac{55\,y^3}{12} \right. \nonumber \\
&& \left. \,+\,(3\,-\,2\,y\,+\,4\,y^2\,-2\,y^3)\log{(1\,-\,y)}
\right)\,,
\eeqa
with $y\,=\,2E_\gamma/m_\mu$.

The gamma-ray fragmentation function from bottom quarks in the dark matter annihilation final state is simulated using {\tt PYTHIA} and  fitted using the following analytic function:
\beqa
\frac{dN^b_\gamma(E)}{dE_\gamma}&=&\frac{2}{E}\,e^{7.59\,-\,80.25\,x\,+\,412.5\,x^2\,-\,1297.7\,x^3} \nonumber \\
&&\quad \times e^{\,+\,1969.7\,x^4\,-\,1137.8\,x^5}\,,
\eeqa
with $x\equiv 2E_\gamma/E$ and  $E$ as the center of mass energy of the $b\,\bar{b}$ system.

Summing up all contributions to the gamma-rays, we compare the model predictions with the background for gamma-ray energy above 1~GeV, which is fitted by a power-law in Ref.~\cite{Bergstrom:1997fj} as 
\beqa
\frac{d^2\Phi^{\rm BG}_\gamma}{d\Omega\,dE_\gamma}\,=\,6\,\times\,10^{-5}\,\left(\frac{E_\gamma}{\rm 1 GeV}\right)^{-2.72}\,,
\eeqa
in ${\rm cm^{-2}s^{-1}sr^{-1} GeV^{-1}}$. The Fermi LAT collaboration has already shown a preliminary result  for the gamma rays from $0^\circ\le l \le 360^\circ$ and $10^\circ\le |b| \le 20^\circ$~\cite{fermiLATGiglietto}. %
\begin{figure}[ht!]
\centerline{ \hspace*{0cm}
\includegraphics[width=0.50\textwidth]{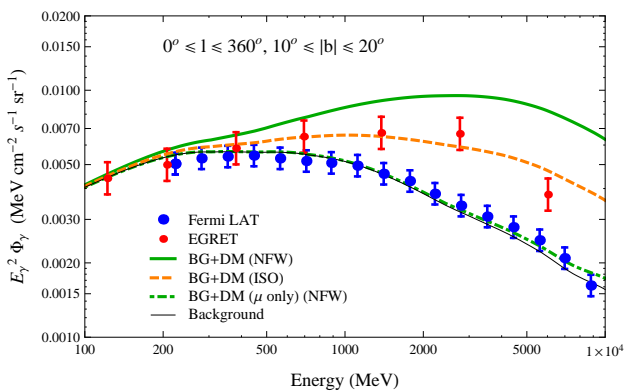}
} 
\caption{The gamma ray flux predicted from the neutralino annihilations in the NMSSM.  
The neutralino mass is $m_\chi\,=\,162$~GeV and its annihilation cross section 
is  $6.0\times 10^{-24}$ cm$^2$ s$^{-1}$. The solid green line is the dark matter 
prediction using the NFW profile plus the traditional background (shown as the thin black line). 
The dashed orange line is the dark matter prediction using the cored isothermal dark matter profile. 
The dot-dashed green line is the dark matter prediction without including the contributions from $b$ quarks. }
\label{fig:fermiLargeAngle}
\end{figure}
Here, $l$ and $b$ are the heliocentric galactic coordinates. This region of angles corresponds to $\Delta \Omega = 0.567$~sr. 
It is easy to calculate $\bar{J} \Delta \Omega$ to be 13.4 for the NFW profile and 5.7 for the ISO profile.
In Fig.~\ref{fig:fermiLargeAngle}, we show the observed gamma ray fluxes from Fermi LAT together 
with the data from EGRET. As one can see, there is a disagreement between those two experiments for photon energy above 1~GeV. 
The predictions from dark matter annihilation using the model point in Table~\ref{tab:case1para} tend to agree with the EGRET result 
if one uses the ISO dark matter profile. 

A larger discrepancy occurs for the dark matter prediction using the NFW profile. 
This might be reconciled by astrophysical uncertainties 
like the smoothness of the dark matter distribution, 
which affects the necessary dark matter annihilation 
cross section by a factor of a few. Notice also
in Fig.~\ref{fig:fermiLargeAngle} the dot-dashed green line showing
the signal gamma-ray contribution from the muon final states only;
this indicates that the discrepancy is coming from the 
$b$ quarks in the final state.

\begin{figure}[ht!]
\centerline{ \hspace*{0cm}
\includegraphics[width=0.50\textwidth]{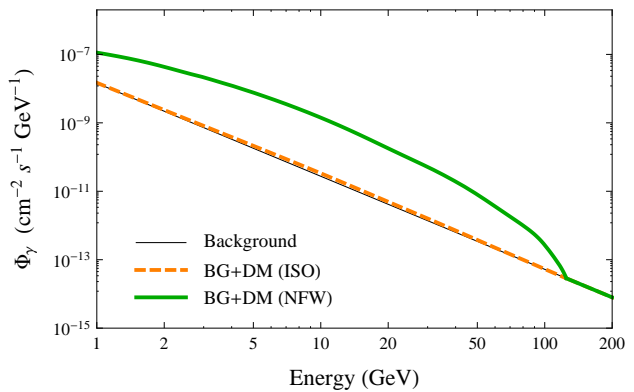}
} 
\caption{The photon flux predicted from the neutralino annihilations in the NMSSM after fitting the positron fraction spectrum of PAMELA.  
The neutralino mass is $m_\chi\,=\,162$~GeV and its annihilation cross section 
is  $6.0\times 10^{-24}$ cm$^2$ s$^{-1}$. $\Delta \Omega=2.4 
\times 10^{-4}$ ($0.5^\circ \times 0.5^\circ$ about the galactic center) 
and $\bar{J}\Delta \Omega \approx 0.69$ (NFW) and $\bar{J}\Delta \Omega \approx 0.003$ (ISO) are used here.}
\label{fig:fermi}
\end{figure}

We can also consider gamma rays coming from the galactic center. 
In Fig.~\ref{fig:fermi}, we compare the differential gamma-ray flux 
predicted from the neutralino annihilations in the NMSSM to the flux from the background, after fitting the positron fraction spectrum of PAMELA. 
We present the comparison of the model predictions and the background for a $0.5^\circ \times 0.5^\circ$ region about the 
galactic center. Since the Fermi LAT will have an angular resolution of around $0.1^\circ \times 0.1^\circ$ ($10^{-5}$~sr), 
the error of their measurements in principle is small enough to test the dark matter annihilation scenario. However, 
the model predictions for the gamma-ray flux in the galactic center are subject to large uncertainties from the dark matter profile, 
which can bring a factor of few hundred difference as seen from Fig.~\ref{fig:fermi}.

Due to kinematical reasons, our dark matter neutralino mass is less than the top quark mass. 
Therefore, the gamma rays predicted in this model have an energy cutoff below the top quark mass. 
Hence, this model automatically evades  the constraints from HESS, which measured the gamma-ray flux 
with energy above 200~GeV. For EGRET with $\Delta \Omega =10^{-3}$ around the galactic center and with 
the energy range of $1 {\rm~GeV}\le E_\gamma\le 30$~GeV, the total background flux is $3\times 10^{-8}$ cm$^{-2}$ s$^{-1}$. 
The dark matter contribution in this model is around $2\times 10^{-7}$ cm$^{-2}$ s$^{-1}$ for the NFW profile 
and $1.5\times 10^{-9}$ cm$^{-2}$ s$^{-1}$ for the ISO profile. This indicates a tension between the model 
prediction with the NFW profile and the EGRET data. Again this result is subject to astrophysical uncertainties; 
for a solid angle around the galactic center, changing the dark matter profile can introduce an uncertainty of two orders 
of magnitude in the gamma-rays flux predictions. 

\section{Direct constraints}
\label{sec:constraints}
There are many direct constraints on the NMSSM parameter space from LEP, Tevatron, CLEO, B-factories and the 
magnetic moment of the muon. Our analysis shows that the most stringent constraints are from exclusive $\Upsilon(3S)$ decays at BaBar
and light pseudoscalar searches at D0. Compared to the MSSM, the constraints from LEP and other searches 
are less severe, due to non-standard decays of the MSSM-like Higgs boson and a new suppression 
factor $\cos{\theta_A}$ beyond the MSSM.

\subsection{LEP constraints}
\label{sec:LEP}
The heavier bosons in the NMSSM, typically above 250~GeV, were
unaccessible at LEP. Therefore, we only consider the two lightest scalar particles $a_1$ and $h_1$ at LEP. There are three main 
production mechanisms for those two neutral Higgs bosons. One is through the Higgsstrahlung process $e^+\, e^-\,\rightarrow\,h_1\,Z$; 
another one is the pair production process $e^+\, e^-\,\rightarrow\,h_1\,a_1$; the third one is through the radiation off a massive 
fermion: $e^+\, e^-\,\rightarrow\,b\,\bar{b}\,a_1$.

For the Higgsstrahlung process $e^+\, e^-\,\rightarrow\,h_1\,Z$ and for an $h_1$ mass within the LEP reach, the $h_1$ mainly decays 
to $b\,\bar{b}$ and $2\,a_1$ for the $\mu$-favored and $\tau$-favored cases, respectively.  For the $\tau$-favored model point, 
there are many different final states like $4\tau$, $2\tau\,2g$ and so on. Although the LEP bounds on each of those channels are 
weak, the decay mode independent limits impose a bound  on the mass of $h_1$ as $M_{h_1}>82$~GeV~\cite{Abbiendi:2002qp}, which is 
satisfied in the model points we have considered in this paper. For the $\mu$-favored case, the lower bound on the $h_1$ mass is 
around 114~GeV, which is also satisfied for the model point reported in  Table~\ref{tab:case1para}.

The cross section of the pair production process $e^+\, e^-\,\rightarrow\,h_1\,a_1$ is proportional 
to $\cos^2{\theta_A}\,M^4_Z/M^4_{a_2}$ from Eq.~(\ref{eq:a1h1Z}), and hence is tiny for $\cos{\theta_A}\,<\,0.3$ 
and $M_{a_2}\,>\,300$~GeV.   For $\sqrt{s}=200$~GeV, the cross section is calculated to be $2\times 10^{-3}$~fb for 
the model parameters in Table~\ref{tab:case1para}. Considering the integrated luminosity of LEP is below 1~fb$^{-1}$, 
there are no constraints on the model parameters from this channel. 

Finally, for the associated production with bottom quarks, the cross section is also suppressed due to a moderate $\tan{\beta}$ 
and three-body final state phase space. The cross section for $e^+\, e^-\,\rightarrow\,b\,\bar{b}\,a_1$ at  $\sqrt{s}=200$~GeV 
is calculated to be $1.5\times 10^{-3}$~fb for the model parameters in Table~\ref{tab:case1para}, which also indicates no constraints from LEP. 

\subsection{Tevatron constraints}
\label{sec:TEVATRON}
The lightest $CP$-even Higgs boson $h_1$ has approximately the same couplings to fermions as in the Standard Model. 
The ongoing searches at CDF and D0 do not yet 
constrain an $h_1$ with a mass around 115~GeV at the Tevatron~\cite{Phenomena:2009pt}.

For the lightest $CP$-odd Higgs boson $a_1$, the main production  process at the Tevatron is through associated production with $b\,\bar{b}$ 
and has a cross section proportional to $\tan^2{\beta}\,\cos^2{\theta_A}\,\sigma (b\,\bar{b}\,\phi_{\rm SM})$, where $\phi$ has the same coupling 
to $b\,\bar{b}$ as the SM Higgs boson. Although the production cross section can be large for the model parameters 
that we consider, the existing searches at D0 for the $a_1$ decays to two taus only constrain an
$a_1$ with a mass above 90~GeV~\cite{Abazov:2008zz}.  

However, the recent searches for $a_1$ in the channel $h_1\,\rightarrow\,a_1a_1\,\rightarrow \mu^+\mu^-\mu^+\mu^-$  at D0 can impose a 
stringent bound on the muon-favored model parameter space in the NMSSM. The SM background for two pairs of collinear muons
is very small (below 0.02 events for $3.7$~fb$^{-1}$ integrated luminosity), and the null result imposes a constraint 
$\sigma(p\bar{p}\rightarrow h_1 +X)\cdot{\rm Br}(h_1\rightarrow a_1a_1)\cdot {\rm Br}(a_1\rightarrow \mu^+\mu^-)^2 \lsim 10$~fb~\cite{D04muon}. 
This bound is roughly independent of the $a_1$ mass. The production cross section for a SM Higgs at the Tevatron is 
around 0.8~pb, so we need to have ${\rm Br}(h_1\rightarrow a_1a_1)\lsim 1.2\%$ for ${\rm Br}(a_1\rightarrow \mu^+\mu^-)\approx 100\%$. 
Therefore, the muon-favored model point in Table~\ref{tab:case1para} is allowed by this constraint, although it selects a specific region of the NMSSM parameter space. This experimental result points to small values of $\lambda$ and $\kappa$ to 
decrease the $h_1$ branching ratio into $a_1a_1$. However, small values of $\lambda$ and $\kappa$ also increase the branching ratio of $a_2$ into $b\,\bar{b}$. Since the dominant annihilation is mediated by the  resonance effect with $a_2$, this leads to a 
non-neglegible hadronic final state from the dark matter annihilation.

\subsection{The magnetic moment of the muon}
\label{sec:muonmag}
There is a new radiative contribution to  the magnetic moment of the muon by exchanging $a_1$ in the loop diagram. 
Using the one-loop result from Ref.~\cite{Krawczyk:2001pe}, the new physics correction to $a_\mu$ is
\beqa
\delta a_\mu&=&\frac{g_2^2\,m_\mu^2}{32\,\pi^2\,M_W^2}\,(\cos{\theta_A}\,\tan{\beta})^2
\nonumber \\
&\times&\frac{m^2_\mu}{M^2_{a_1}}\,\int^1_0\,\frac{-x^3\,dx}{x^2(m^2_\mu/M^2_{a_1})\,+\,1\,-\,x}\,,
\eeqa
and is negative. The two-loop calculation will not change the sign of $\delta a_\mu$ for a mass of $a_1$ below 1~GeV~\cite{Gunion:2005rw}.  
The measured value of $a_\mu$ has a 3.4 $\sigma$
deviation ($e^+\,e^-$ data only) above the prediction of the standard model~\cite{Bennett:2006fi}. Requiring the new physics to be less than the 
experimental error, we arrive at the following constraints on the model parameters: 
\beqa
\cos{\theta_A}\,\tan{\beta}\,\lsim\,2.5\qquad ({{\rm BNL}\;\; (g-2)_\mu/2})\,.
\eeqa
Here $M_{a_1}$ is chosen to be 800~MeV, while the constraints are less stringent as one increases $M_{a_1}$. 

Notice that the $\delta a_\mu$ may also receive significant contributions from 
other particles like smuons, that we chose to be heavy here to isolate the dark matter discussions.

\subsection{Constraints from Upsilon decays}
\label{sec:upsilon}
Another stringent bound on the NMSSM parameter space with a light
$a_1$ below 10 GeV is from Upsilon decays 
into a photon plus $a_1$, which decays into a pair of taus or muons.

For the mass range $2m_\tau\,<\,m_{a_1}\,<\,9.2$~GeV, the strongest bound is 
from the recent CLEO-III limits~\cite{:2008hs} on $\Upsilon(1S)\,\rightarrow\,\gamma\,\tau^+\tau^-$. 
The radiative decay to $\Upsilon(1S)\,\rightarrow\,\gamma\,a_1$ is calculated as~\cite{Wilczek:1977zn}
\beqa
\frac{{\cal B}(\Upsilon(1S)\,\rightarrow\,\gamma\,a_1)}{{\cal B}(\Upsilon(1S)\,\rightarrow\,\mu^+\,\mu^-)}&=&\frac{G_F\,m_b^2}{\sqrt{2}\,\pi\,\alpha}\,(\cos{\theta_A}\,\tan{\beta})^2\,
\nonumber \\
&\times&\left(1\,-\,\frac{M_{a_1}^2}{M_{\Upsilon(1S)}^2}\right)\,{\cal F}\,,
\eeqa
where ${\cal F} \sim 0.5$  incorporates QCD and relativistic corrections~\cite{Mangano:2007gi}~\cite{Aznaurian:1986hi}. 
The data from CLEO have the limit  ${\cal B}(\Upsilon(1S)\,\rightarrow\,\gamma\,a_1)\times {\cal B}(a_1\,\rightarrow\,\tau^+\,\tau^-)\,\lesssim\,5\times 10^{-5}$ at 90\% C.L. for a wide range of $M_{a_1}$ between $4$~GeV to $9$~GeV. 
Using ${\cal B}(\Upsilon(1S)\,\rightarrow\,\mu^+\,\mu^-)\,=\,2.48\%$ 
and ${\cal B}(a_1\,\rightarrow\,\tau^+\,\tau^-)\,\approx\,0.9$ from the model prediction, this limit 
is translated into a bound on $\cos{\theta_A}\,\tan{\beta}$ as
\beqa
\cos{\theta_A}\,\tan{\beta}\,\lesssim\,0.9\qquad {\rm (CLEO-III)}\,.
\eeqa
A similar result is obtained in~\cite{Domingo:2008rr}. For the mass range $2m_\mu\,<\,m_{a_1}\,<1$~GeV, the strongest current bound is coming from the light scalar searches in the channel $\Upsilon(3S)\,\rightarrow\,\gamma\,a_1$ at BaBar. At 90\% C.L., BaBar imposes an upper 
limit ${\cal B}(\Upsilon(3S)\,\rightarrow\,\gamma\,a_1)\times {\cal B}(a_1\,\rightarrow\,\mu^+\,\mu^-)\,\lsim\,5.2\times 10^{-6}$~\cite{Aubert:2009cp} for $M_{a_1}$ below 1~GeV. Using ${\cal B}(\Upsilon(3S)\,\rightarrow\,\mu^+\,\mu^-)\,=\,2.18\%$ and ${\cal B}(a_1\,\rightarrow\,\mu^+\,\mu^-)\,\approx\,1.0$ from the model prediction, this limit is  translated into a bound on $\cos{\theta_A}\,\tan{\beta}$ as
\beqa
\cos{\theta_A}\,\tan{\beta}\,\lesssim\,0.4\qquad {\rm (BaBar)}.
\label{eq:babar}
\eeqa

For the range of  $2m_\mu<m_{a_1}<M_K-M_\pi$, the decay mode $K^+\,\rightarrow\,\pi^+\,a_1$ is open. 
The branching ratio of this decay channel is given by~\cite{Bardeen:1978nq}
\beqa
&&{\cal B}(K^+\,\rightarrow\,\pi^+\,a_1)=\frac{G_F\,f_\pi^2}{\sqrt{2}}\,(\tan{\beta}\,-\,\tan^{-1}{\beta})^2\, \nonumber \\
&&\hspace{2.5cm} \times \cos^2{\theta_A}\,\frac{\Gamma(K^0_s\,\rightarrow\,\pi^0\,\pi^0)}{\Gamma(K^+\,\rightarrow\,{\rm all})}   \nonumber \\
&=& 3\times 10^{-6}\,(\tan{\beta}\,-\,\tan^{-1}{\beta})^2\,\cos^2{\theta_A}\,,
\eeqa
which should be compared to the experimental values from the 
HyperCP collaboration~\cite{Park:2001cv}: ${\cal B}(K^+\,\rightarrow\,\pi^+\,\mu^+\,\mu^-)\,=\,9.8 \pm 1.0 \pm 0.5\times 10^{-8}$. 
Therefore, the following constraint on the model parameter space is derived  
\beqa
\cos{\theta_A}\,|\tan{\beta}\,-\,\tan^{-1}{\beta}|\,\lesssim\,0.06\qquad {\rm (HyperCP)}\,,
\eeqa
So, other than when $\tan{\beta}$ is very close to 1 or $\cos{\theta_A}$ is extremely close to zero, the mass of $a_1$ is constrained to be above $M_K-M_\pi \sim 360$~MeV.

Finally, there are also other constraints from B-physics like $b\,\rightarrow\,s\,\gamma$ or $B_s\rightarrow \mu^+\mu^-$. 
Since the $a_1$ pseudoscalar does not mediate tree-level flavor-changing processes, one can use the minimal flavor violation assumption to 
suppress many kinds of flavor changing processes. We have used the NMHDECAY program to check those constraints and to justify the validity of our model points. In short, the most stringent constraint for $M_K-M_\pi< m_{a_1} < 1$~GeV is the radiative decays of Upsilon into photons from BaBar. The bound is $\cos{\theta_A}\,\tan{\beta}\,\lesssim\,0.4$. One can see that the  model point in Table~\ref{tab:case1para} satisfies this bound. 

\section{Discussions and Conclusions}
\label{sec:conclusions}
The dark matter candidate neutralino  in the NMSSM from Table~\ref{tab:case1para} 
is a combination of the bino, wino and Higgsino. Therefore, it has a good chance 
to be detected in dark matter direct detection experiments, especially from 
the spin-dependent elastic scattering with nucleons. Here we
just report the values calculated using the micrOMEGAs program~\cite{Belanger:2008sj}. 
The spin-independent DM-proton and DM-neutron cross sections are $0.7\times 10^{-45}$~cm$^2$ and $0.9\times 10^{-45}$~cm$^2$, which are two orders of magnitude below the current bounds from XENON10~\cite{Angle:2007uj}. 
For the spin-dependent one, the DM-proton and DM-neutron cross sections are $1.5\times 10^{-39}$~cm$^2$ and $1.2 \times  10^{-39}$~cm$^2$. 
The later one is only one order of magnitude below the current bound~\cite{Angle:2008we} and is in the accessible region of the upgraded experiments. 

The NMSSM explanation of the PAMELA positron excess can be tested by future experiments at colliders and new results 
from cosmic ray experiments. On the collider side, it is important to measure the masses of the neutralino and the heavier $CP$-odd particle. 
If their masses satisfy the relation $M_{a_2}\,\approx\,2\,m_\chi$, the large dark matter annihilation cross section can be confirmed. 
Another important quantity to measure is the mass of the lighter $CP$-odd particle, because the PAMELA positron excess prefers to have 
its mass below 1~GeV. It is also crucial to know the branching ratio of $a_2$ to $b\,\bar{b}$, since the dominant antiproton contributions 
are from this channel. 

On the cosmic ray side, we make a well-motivated unambiguous prediction that PAMELA will observe a turnover of the rising positron spectrum at around 70 GeV. The additional contributions to the electron+positron spectrum from dark matter annihilations are within current uncertainties but could be resolved in the future by Fermi LAT. The associated gamma-ray flux from dark matter annihilation could also be resolved by Fermi LAT, but again due to astrophysical uncertainties one should be cautious when making a concrete prediction. 
    
In this paper we have explored the possibility of using neutralino annihilations in the NMSSM to explain the positron excess observed at PAMELA. Kinematics plays an essential role for having a viable model with a large fraction of leptons and a small fraction of hadrons in the annihilation final state. The lighter $CP$-odd particle $a_1$ has a mass below 1~GeV and mainly decays into two muons. The dark matter neutralino mass is less than the top quark mass to forbid the otherwise dominant $t\,\bar{t}$ final state. The $s$-channel resonance effect with the heavier $CP$-odd particle $a_2$ increases the dark matter annihilation cross section to match the necessary one for the PAMELA positron fraction spectrum. This also requires that the neutralino mass is less than the top quark mass, to avoid smearing the resonance effect from a
large $t\,\bar{t}$ contribution to the $a_2$ width. 

We have also shown that there is discrepancy between the NMSSM predictions and the preliminary gamma ray fluxes from Fermi LAT. To alleviate this descrepancy, one could evoke the existence of an astrophysical boost factor from a nearby clump of dark matter. Such boost factor would also improve the agreement between the NMSSM prediction for the antiproton spectrum and the PAMELA results. In particular, given the fact that positrons/electrons are only coming from nearby sources, a clump of dark matter can affect positron/electron, antiproton/proton and gamma ray fluxes differently, and would relatively increase the positron ratio spectrum more than the antiproton ratio spectrum and the gamma ray flux spectrum. We have not utilized this astrophysical boost factor in our analysis.

An NMSSM explanation of PAMELA makes three striking and uniquely correlated predictions: the rise in the PAMELA positron spectrum will turn over at around 70~GeV, 
the dark matter particle mass is less than the top quark mass, and a light sub-GeV pseudoscalar will be discovered at colliders.

\bigskip

{\bf Acknowledgments:} 
The authors are grateful to Laura Covi, Gordon Kane, Maurizio Pierini and Peter Skands for useful discussions. Fermilab is operated by Fermi Research Alliance, LLC under contract no. DE-AC02-07CH11359 with the United States Department of Energy.  

\vspace{2mm}

{\it{Note: After this paper was submitted in the arXiv, the BaBar collaboration presented a new constraint on the Upsilon radiative decays to photon plus two muons by combining both $\Upsilon(2S)$ and $\Upsilon(3S)$ data~\cite{Aubert:2009cp2}.  This implies a more stringent bound on $\cos{\theta_A}\tan{\beta}$  by a factor of two, see Eq.~(\ref{eq:babar}). This can be accommodated in the NMSSM by increasing  the branching ratio of $a_2$ to $b\,\bar{b}$. This increases the tension for the antiproton ratio spectrum observed at PAMELA. However, a modest astrophysical boost factor on the order of $\lesssim5$ would loosen this tension.}}

 
\vfil \end{document}